\documentclass[aps,prd,twocolumn,superscriptaddress,amssymb,amsmath,nofootinbib,showpacs,balancelastpage,10pt]{revtex4-1}
\usepackage{graphicx,longtable,natbib,mathrsfs,color,bm}

\newcommand{\de}{\mathrm d}

\newcommand{\om}{{\Omega_m}}

\newcommand{\orad}{{\Omega_r}}

\newcommand{\ho}{{H_0}}

\newcommand{\lcdm}{$\Lambda$CDM}

\begin{document}

\addtolength{\hoffset}{-0.165cm}
\addtolength{\textwidth}{0.33cm}
\title{Detectability of Torsion Gravity via Galaxy Clustering and Cosmic Shear Measurements}

\author{Stefano Camera}
\email{stefano.camera@tecnico.ulisboa.pt}
\affiliation{CENTRA, Instituto Superior T\'ecnico, Universidade de Lisboa,
Av. Rovisco Pais 1, 1049-001 Lisboa, Portugal}
\author{Vincenzo F. Cardone}
\affiliation{INAF, Osservatorio Astronomico di Roma, Via Frascati 33, 00040 Monte Porzio Catone (Roma), Italy}
\author{Ninfa Radicella}
\affiliation{Dipartimento di Fisica ``E.R. Caianiello'', Universit\`a di Salerno, and INFN, Sez. di Napoli, GC di Salerno, Via Giovanni Paolo II 132, 84084 Fisciano (SA), Italy}

\begin{abstract}
Alterations of the gravity Lagrangian introduced in modified torsion gravity theories---also referred to as $f(T)$ gravity---allows for an accelerated expansion in a matter dominated Universe. In this framework, the cosmic speed up is driven by an effective `torsion fluid'. Besides the background evolution of the Universe, structure formation is also modified because of a time dependent effective gravitational constant. Here, we investigate the imprints of $f(T)$ gravity on galaxy clustering and weak gravitational lensing to the aim of understanding whether future galaxy surveys could constrain torsion gravity and discriminate amongst it and standard general relativity. Specifically, we compute Fisher matrix forecasts for two viable $f(T)$ models to both infer the accuracy on the measurement of the model parameters and evaluate the power that a combined clustering and shear analysis will have as a tool for model selection. We find that with such a combination of probes it will indeed be possible to tightly 
constrain $f(T)$ model parameters. Moreover, the Occam's razor provided by the Bayes factor will allow us to confirm an $f(T)$ power-law extension of the concordance \lcdm\ model, were a value larger than $0.02$ of its power-law slope measured, whereas in \lcdm\ it is exactly $0$.
\end{abstract}


\maketitle

\section{Introduction}\label{sec:introduction}
The accelerated cosmic expansion has been confirmed up to now by a wide range of cosmological datasets, from type Ia supernov\ae\ \citep[SNeIa;][]{Riess:1998cb,Perlmutter:1998np}, to the cosmic microwave background (CMB) radiation \citep{Hinshaw:2012aka}, baryon acoustic oscillations \citep[BAOs;][]{Eisenstein:2005su} and the gamma ray burst (GRB) Hubble diagram \citep{Wang:2008vja}. Although these pieces of evidence can fit the framework of general relativity (GR) if we assume the presence of a cosmological constant term in Einstein's field equations, this is deeply unsatisfactory an answer, from a theoretical viewpoint \citep[e.g.][]{Weinberg:1988cp}. Conversely, the idea that we may instead be disregarding some gravitational effect occurring on cosmological scales is rather intriguing, and somehow follows an Einstein inspired approach---i.e.\ to look for a generalisation of the law of gravity whereby data requires it.

Amongst the wide class of the extended theories of gravity, we here consider the so-called $f(T)$ gravity theory. It is a generalisation of the teleparallel gravity, where torsion, instead of curvature, is responsible for the gravitational interaction \cite{Einstein:28,Einstein:30a, Einstein:30b}. As a consequence, the torsion scalar $T$ replaces the curvature scalar $R$ in the Lagrangian. In this framework, the underlying Riemann-Cartan spacetime is endowed with the Weitzenbock connection, that is curvature free. In this scenario torsion acts as a force, allowing for the interpretation of gravity as a gauge theory of the translation group \citep{Arcos:2005ec}. Despite conceptual differences, teleparallel gravity and GR yield thoroughly equivalent dynamics, the interpretation of the gravitational interaction in terms of a spacetime with curvature or torsion being therefore only a matter of convenience, at least at the classical level.

Nevertheless, when one generalises teleparallel gravity to a modified $f(T)$ version, inspired by the $f(R)$ extended gravity theories \cite{Capozziello:2003tk,Carroll:2003wy}, the equivalence with GR breaks down: the two classes of models differ in facts \citep{Ferraro:2008ey, Fiorini:2009ux}. Differently from $f(R)$ theories, that can be viewed as a low-energy limit of some fundamental theory, $f(T)$ gravity is just a phenomenological extension of teleparallelism but preserves the advantage of giving equations that are still second order in field derivatives, oppositely to the fourth order equations deduced in $f(R)$ gravity.  Then, it would be interesting to test it as a possible alternative candidate for a theory providing an accelerated cosmic expansion without the need of any exotic component. However, there is a caveat, as these models suffer from the lack of local Lorentz invariance. It means that all the $16$ components of the vierbien are independent and one cannot simply get rid of $6$ of them by fixing a specific gauge \citep{Li:2010cg}.

Moreover, we want to emphasise that $f(T)$ gravity does not belong to the vast family of models reproduced by the Horndeski Lagrangian, that actually includes  scalar-field dark energy models \citep{2010deto.book.....A}, but also modified gravity theories such as $f(R)$ and $f(\mathcal G)$ gravity \citep{DeFelice:2010aj,Nojiri:2005vv,Koivisto:2006xf}, scalar-tensor (including Brans-Dicke) models \citep{Brans:1961sx,EspositoFarese:2000ij}, $K$-essence \citep{ArmendarizPicon:1999rj,ArmendarizPicon:2000ah}, and Galileons \citep{Nicolis:2008in,Burrage:2011bt,Gao:2011mz}. Then, it is worth scrutinising generalised torsion cosmologies, since they cannot be confirmed or ruled out on the basis of an analysis performed for Horndeski models.

Motivated by these considerations, in Reference~\citep{Cardone:2012xq} we have analysed two $f(T)$ gravity models that present the interesting feature of an effective equation of state parameter, $w_\mathrm{eff}(z)$, crossing the so-called phantom divide line, i.e.\ $w_\mathrm{eff}=-1$. We have showed that both models are in very good agreement with a wide set of data, including SNIa and GRB Hubble diagrams, BAOs at different redshifts, Hubble expansion rate measurements and the WMAP7 distance priors. Yet, that wide dataset is unable to severely constrain the model parameters and hence discriminate amongst the considered $f(T)$ models and the \lcdm\ scenario. The point is that the data only probe the Universe's background expansion history. Here, we present a step forward, focussing on the sub-horizon limit, where torsion gravity leads to a rescaling of Newton gravitational constant by a time-dependent factor that explicitly depends upon the modified Lagrangian. As a consequence, the growth of perturbations 
is different compared to what predicted in the \lcdm\ model, and can be tested by present and oncoming surveys designed to probe the large-scale cosmic structure, such as the Dark Energy Survey\footnote{\texttt{http://www.darkenergysurvey.org}} \citep[DES;][]{Abbott:2005bi}, \textit{Euclid}\footnote{\texttt{http://www.euclid-ec.org}} \citep{EditorialTeam:2011mu,Amendola:2012ys}, Pan-STARRS\footnote{\texttt{http://pan-starrs.ifa.hawaii.edu}} or the Large Synoptic Survey Telescope\footnote{\texttt{http://www.lsst.org}} \citep[LSST;][]{Ivezic:2008fe,Abate:2012za} in the optical and near infrared bands, or the Square Kilometre Array\footnote{\texttt{http://www.skatelescope.org}} \citep[SKA;][]{2013IAUS..291..337T} and its pathfinders \citep[e.g.][]{2008ExA....22..151J,Oosterloo:2010wz,Norris:2011ai,Rottgering:2011jq} in the radio band.

We calculate both the three-dimensional and the projected, two-dimensional matter power spectrum from galaxy clustering and the cosmic shear signal, as predicted in viable $f(T)$ cosmologies. Then, we study the constraining potentiality of an \textit{Euclid}-like survey. First, we focus on one of the models already tested in Ref.~\citep{Cardone:2012xq}, which not only successfully passes geometrical data tests, but also shows agreement with growth data. Secondly, we analyse the case where the Universe is correctly described by \lcdm, but the true, underlying cosmology is actually an $f(T)$ model whose parameter space `contains' that of \lcdm. In this case, we also ask ourselves for which of the competing theoretical frameworks is preferred, given the data. We do so by calculating the Bayes factor \citep{Trotta:2005ar,Heavens:2007ka,Camera:2011ms,Camera:2010wm}, in the context of the model selection problem.

The layout of the paper is as follows. A summary of the main equations for $f(T)$ theories is given in Section~\ref{sec:fT-gravity}, where we also present the models we will investigate. The observational probes used as input to the Fisher matrix forecasts are discussed in Sect.~\ref{sec:observables}, whilst the results obtained when using each single probe separately or in combination are given in Sect.~\ref{sec:results}. Bayesian model selection is discussed in Sect.~\ref{sec:lnB}. A summary and future perspectives are finally given in Sect.~\ref{sec:conclusions}.

\section{Modified Torsion Gravity}\label{sec:fT-gravity}
Teleparallelism promotes the vierbein field $e^{a}_{\mu}(x)$ to the r\^ole of a dynamical object with components related to the metric tensor, as
\begin{equation}
g_{\mu \nu}(x) = \eta_{a b} e^a_\mu(x) e^b_\nu(x) \ ,
\end{equation}
where $\eta_{a b} = \text{diag}(1,-1,-1,-1)$. Notice that Latin indices refer to the tangent space whilst Greek letters label coordinates on the manifold. The dynamics is then described by the Lagrangian
\begin{equation}
\mathcal L=\frac{e}{16 \pi G}\left[T+f(T)\right]+\mathcal L_M,\label{eq:lagrangian}
\end{equation}
where $e\equiv\det e^a_\mu=\sqrt{-\det g_{\mu\nu}}$, $\mathcal L_M$ is the matter field Lagrangian and the term $f(T)$ originates the deviations from standard GR. It is a generic function of the torsion scalar $T$ which is defined as
\begin{equation}
T=\frac{1}{4}T^{\lambda\mu\nu}T_{\lambda\mu\nu}+\frac{1}{2}T^{\lambda\mu\nu}T_{\nu\mu\lambda}-T_{\mu\nu}^{\ \ \mu} T^{\lambda\nu}_{\ \ \lambda},
\end{equation}
with the torsion tensor given by
\begin{equation}
T^\lambda_{\mu \nu} = e^\lambda_a \left( \partial_\nu e^a_\mu - \partial_\mu e^a_\nu \right).
\end{equation}

By varying the action with respect to the vierbein $e^a_\mu(x)$, one gets the field equations
\begin{multline}
e^{-1}\partial_\mu(e\   e_a^\rho S_{\rho}^{\ \mu\nu})(1+f_{,T})+e_{a}^\lambda S_{\rho}^{\ \nu\mu} T^{\rho}_{\ \mu\lambda} (1+f_{,T})\\
+e^{\rho}_a S_{\rho}^{\ \mu\nu}\partial_\mu (T) f_{,TT}+\frac{1}{4}e_a^\nu (T+f) = 4\pi G e_a^\mu \Theta_\mu^\nu,\label{eq:fieldeqs}
\end{multline}
where with $\Theta^\nu_\mu$ we indicate the matter energy-momentum tensor, not to create ambiguities with the torsion tensor; here, a comma denotes a derivative with respect to $T$.

To investigate cosmology, it should be kept in mind that two pair of vierbein that lead to the same metric tensor are not equivalent from the point of view of the theory. It means that we are not allowed to simply insert the Friedmann-Lema\^itre-Robertson-Walker (FLRW) metric into Eqs~\eqref{eq:fieldeqs}. Nevertheless, in case of spatially flat metric, a convenient choice is represented by the diagonal vierbein \citep{Linder:2010py,Wu:2010mn,Bengochea:2010sg,Ferraro:2011us}, i.e.\
\begin{align}
e^0&=dt,\\
e^i&=a(t)dx^i,
\end{align}
where $a(t)$ is the scale factor as function of cosmic time $t$. With such a choice, the modified Friedmann equations become
\begin{align}
H^2&=\frac{8\pi G}{3}\rho-\frac{1}{6}f(T)-2H^2f_{,T}(T)\nonumber\\
\left(H^2\right)^{\prime}&=\frac{16\pi Gp+6H^2+f(T)+12H^2f_{,T}(T)}{24H^2f_{,TT}(T)-2-2f_{,T}(T)},\label{eq:ftfried}
\end{align}
with $H=\de\ln a/\de t$ the usual Hubble parameter and $\rho(t)$ and $p(t)$ the (background) energy density and pressure of the matter component, respectively. Note that hereafter we will denote with a prime and with a dot differentiation with respect to $\ln a$ and $t$, respectively. In this case, the torsion scalar reduces to $T=-6H^2$.

Eqs~\eqref{eq:ftfried} can be rewritten in the usual form by introducing an effective `dark torsion' fluid with energy density $\rho_T$ such as
\begin{equation}
H^2=\frac{8\pi G}{3}\left[\rho_m+\rho_T\right],
\end{equation}
with
\begin{equation}
\rho_T=\frac{2Tf_{,T}(T)-f(T)}{16\pi G}.\label{eq:ftrho}
\end{equation}
Since matter still minimally couples to gravity, its conservation equation will be unaffected so that we still have $\rho_m\propto a^{-3}$ and $\rho_r\propto a^{-4}$ for the scaling laws of dust matter and radiation. Imposing the Bianchi identities, the conservation equation for the effective torsion fluid reads
\begin{equation}
\dot{\rho}_T+3H(1+w_T)\rho_T=0,
\end{equation}
having defined
\begin{equation}
w_T=-\frac{f/T-f_{,T}+2Tf_{,TT}+\orad(f_{,T}+2Tf_{,TT})/3}{(1+f_{,T}+2Tf_{,TT})(f/T-2f_{,T})}\label{eq:fteos}
\end{equation}
the equation-of-state parameter of the dark torsion fluid. Note the coupling to the radiation energy density through the term $\orad(a)=8\pi G\rho_r(a)/3H^2(a)$. For $f(T)=0$, one has $\rho_T=0$ and modified teleparallel gravity goes back to the standard GR, while the choice $f(T)=\mathrm{const.}$ gives $w_T=-1$ and the \lcdm\ model is recovered.

Eqs~\eqref{eq:ftrho} and \eqref{eq:fteos} clearly show the key r\^ole played by the choice of the $f(T)$ functional expression in determining the dynamics of the Universe. Here, we shall consider two different models. Motivated by the results in Ref.~\citep{Cardone:2012xq}, as a first case, we set
\begin{equation}
f(T)=\alpha(-T)^{n_T}\left(1-e^{p_TT_0/T}\right),\label{eq:ftexp}
\end{equation}
where a $0$ subscript denotes the present-day value of a quantity, and the constant $\alpha$ may be set as function of $\om$, $\orad$ and the $f(T)$ parameters, $n_T$ and $p_T$, as detailed in Ref.~\citep[][and refs therein]{Cardone:2012xq}. In the following, we will refer to this case as the `exp' $f(T)$ model. Note that, in Ref.~\citep{Cardone:2012xq}, we have also investigated a different model, but we discard it here since it is not in agreement with available measurements of the growth rate.

Although in agreement with data probing the background expansion, the $f(T)$-exp model of Eq.~\eqref{eq:ftexp} does not reduce to \lcdm\ for any particular choice of the $(n_T,\,p_T)$ parameters. Albeit this is an interesting feature in its own, such a peculiarity does not allow us to investigate whether clustering and shear data can discriminate between torsion gravity and GR. Therefore, as a second case, we consider a power-law (hereafter `pl') model given by \cite{Bengochea:2008gz}
\begin{equation}
f(T)=\alpha(-T)^{n_T},\label{eq:ftpl}
\end{equation}
where, again, $\alpha$ may be expressed as a function of $\om$, $\orad$ and $n_T$. Note that the $f(T)$-pl model exactly reduces to \lcdm\ for $n_T=0$. As will see later on, we can take $n_T=0$ as fiducial value and look at how strong are constraints on $n_T$, thus quantifying whether or not clustering and shear data can discriminate between modified torsion gravity and GR. 

For what concerns cosmological perturbations, the same caveat as before should be considered when perturbing the metric and the simplest choice may lead to inconsistencies. That is, focusing on the scalar degrees of freedom, one must perturb the vierbein with 6 unknown functions and then choose the longitudinal gauge on the perturbed metric tensor. This reduces the number of free functions to 3, one degree of freedom more than in the GR case \citep{Zheng:2010am}. Nevertheless, this term  plays an important r\^ole on the evolution of perturbations at large scales. In the subhorizon limit this leads to an effective gravitational constant, with respect to the Newtonian constant $G_N$, which takes the form
\begin{equation}
\mathcal G_\mathrm{eff}(z)=\frac{G_N}{1+f_{,T}[T(z)]},\label{eq:ftgeff}
\end{equation}
so that it can be straightforwardly evaluated once the modified Friedmann equations have been solved.

\section{Cosmological Observables}\label{sec:observables}
We adopt the Fisher matrix formalism \citep{Fisher:1935,Jungman:1995bz,Tegmark:1996bz} to make predictions on the $f(T)$ cosmological models presented in Sect.~\ref{sec:fT-gravity}. By doing so, we can scrutinise to which degree of accuracy one of the future large-scale surveys will be able to constrain $f(T)$ model parameters, thus allowing us to discriminate between it and \lcdm---were the constraints tight enough. In the assumption of a Gaussian likelihood, $\mathcal L$, for the model parameters, $\boldsymbol\vartheta=\{\vartheta_\alpha\}$, the Fisher matrix approximates the inverse of the parameter covariance matrix in a neighbourhood of the likelihood peak, i.e.
\begin{equation}
\mathbf F=-\left\langle\frac{\partial^2\ln\mathcal  L}{\partial\boldsymbol\vartheta^2}\right\rangle;\label{eq:fisherm}
\end{equation}
the marginal error on parameter $\vartheta_\alpha$ is thence $\sigma(\vartheta_\alpha)=\sqrt{\left(\mathbf F^{-1}\right)_{\alpha\alpha}}$. Each of the cosmological probes that we study here will then produce its own Fisher matrix, viz.\ $\mathbf F^{g_\mathrm{3D}}$, $\mathbf F^{g_\mathrm{2D}}$  and $\mathbf F^\gamma$ for three- and two-dimensional galaxy clustering and cosmic shear tomography, respectively. We also introduce another important quantity, namely the correlation between the parameter pair $(\vartheta_\alpha,\,\vartheta_\beta)$, which reads
\begin{equation}
r(\vartheta_\alpha,\vartheta_\beta)=\frac{\left(\mathbf F^{-1}\right)_{\alpha\beta}}{\sqrt{\left(\mathbf F^{-1}\right)_{\alpha\alpha}\left(\mathbf F^{-1}\right)_{\beta\beta}}}.\label{eq:correlation}
\end{equation}
This quantity tells us whether the two parameters are completely uncorrelated, when $r(\vartheta_\alpha,\vartheta_\beta)=0$, or thoroughly degenerate, if $r(\vartheta_\alpha,\vartheta_\beta)=\pm1$.

\subsection{3D Galaxy Clustering}\label{ssec:clustering3d}
Large-scale galaxy redshift surveys allow us to investigate the clustering properties of galaxies through measurements of their correlation function and its Fourier transform---the power spectrum, the observable we consider here. BAOs at the last scattering surface give rise to a characteristic peak at the typical scale of $\sim150\,\mathrm{Mpc}$ in the galaxy correlation function, which translates into wiggles in the matter power spectrum. This scale may be taken as a standard ruler, fixed by the sound horizon at last scattering and accurately measured by CMB experiments. By comparing the BAO peak position at the different redshifts, we can constrain both the Hubble parameter $H(z)$, in the radial direction, and the comoving angular diameter distance $d_A(z)$, perpendicularly to the line of sight. Since the underlying cosmology is not known \textit{a priori}, the distance to an object is hence unknown, and what galaxy surveys actually measure is the clustering in the redshift space. As a consequence, the 
power spectrum also contains the imprint of the linear growth rate of structure in the form of a measurable anisotropy due to the coherent flows of matter from low to high densities. When redshift is used to replace distances, peculiar velocities of galaxies introduce distortions in the clustering pattern which can be observed as anisotropies in the correlation function. At linear order, such redshift space distortions (RSDs) depend upon $g(z)\sigma_8(z)$, where $g[z(a)]=\de\ln D_+/\de\ln a$ is the growth rate, $D_+(z)$ the growth factor, $\sigma_8(z)=\sigma_8(z=0)D_+(z)$ and $\sigma_8$ is the variance of the density perturbations on the scale of $8h^{-1}\,\mathrm{Mpc}$. We can constrain $\sigma_8$ via CMB measurements, RSDs are consequently a powerful probe of $g(z)$, thus being able of discriminating amongst different dark energy models and modified gravity theories.

The Fisher matrix for galaxy clustering as measured from galaxies in a redshift bin centred on $z$ reads \citep{Tegmark:1997rp}
\begin{equation}
\mathbf F^{g_\mathrm{3D}}_{\alpha\beta}=\int_{k_\mathrm{min}}^{k_\mathrm{max}}\!\!\frac{\de^3k}{\left(2 \pi\right)^3}\frac{1}{2}\frac{\partial\ln P_\mathrm{obs}(\mathbf k)}{\partial\vartheta_\alpha}\frac{\partial\ln P_\mathrm{obs}(\mathbf k)}{\partial\vartheta_\beta}V_\mathrm{eff}(\mathbf k),\label{eq:fijgal}
\end{equation}
where $P_\mathrm{obs}(\mathbf k) = P_\mathrm{obs}(k, \mu)$ is the anisotropic observed power spectrum (with $\mu$ the cosine of the angle with the line of sight) and $V_\mathrm{eff}(\mathbf k)$ is the survey effective volume. In each redshift bin, it is given by
\begin{equation}
V_\mathrm{eff}(k,\mu)=V_\mathrm{survey}\left\{1+\left[\frac{\de N^\mathrm{(sp)}}{\de z}P_\mathrm{obs}(k,\mu)\right]^{-1}\right\}^{-2},
\end{equation}
with $V_\mathrm{survey}(z)$ the survey volume probed by galaxies in the redshift bin centred on $z$ and $\de N^\mathrm{(sp)}/\de z(z)$ the number density of galaxies with measured spectroscopic redshift in the redshift interval $[z,\,z+\de z]$. We set $k_\mathrm{min}=0.001\,h\,\mathrm{Mpc^{-1}}$ to safely remain in the sub-horizon limit, and $k_\mathrm{max}=0.15\,h\,\mathrm{Mpc^{-1}}$ to stay in the linear r\'egime. Note that future galaxy surveys will measure $P_\mathrm{obs}(k,\mu)$ up to much larger $k$ values, but we prefer here to avoid such small scales and neglect poorly understood non-linear effects. Indeed, they have yet to be investigated---e.g.\ through $N$-body simulations---in $f(T)$ theories and so a mapping from the linear to the non-linear power spectrum is unavailable at the time being. Setting $k_\mathrm{max}=0.15\,h\,\mathrm{Mpc^{-1}}$ guarantees that we are in the linear r\'egime so that we do not add further uncertainties or systematic error due neglecting or incorrectly modelling non-linearities.

The observed power spectrum is a distorted representation of the underlying matter power spectrum
\begin{equation}
P^\delta(k,z)=A_sk^{n_s}\left[\mathcal T(k)D_+(z)\right]^2,\label{eq:pdd}
\end{equation}
where $\mathcal T(k)$ is the transfer function, which we calculate following Ref.~\citep{Eisenstein:1997ik} fitting formul\ae, $n_s$ is the spectral index and the normalisation constant $A_s$ can be related to $\sigma_8$. The redshift dependence is introduced through the linear growth factor $D_+(z)$ which can be conveniently computed by integrating the growth rate, $g(z)$. In the sub-horizon limit we are interested in here, it may be obtained as the solution of the following non-linear differential equation
\begin{multline}
\frac{\de g(z)}{\de z}+\left[\frac{\de\ln{E^2(z)}}{\de\ln(1 + z)}+2+g(z)\right]\frac{g(z)}{1+z}\\+\frac{3}{2}\frac{\om\left(1+z\right)^2}{E^2(z)}\frac{\mathcal G_\mathrm{eff}(z)}{G_N}=0,\label{eq:greq}
\end{multline}
with $E(z)=H(z)/\ho$.

\begin{figure}
\centering
\includegraphics[width=0.5\textwidth]{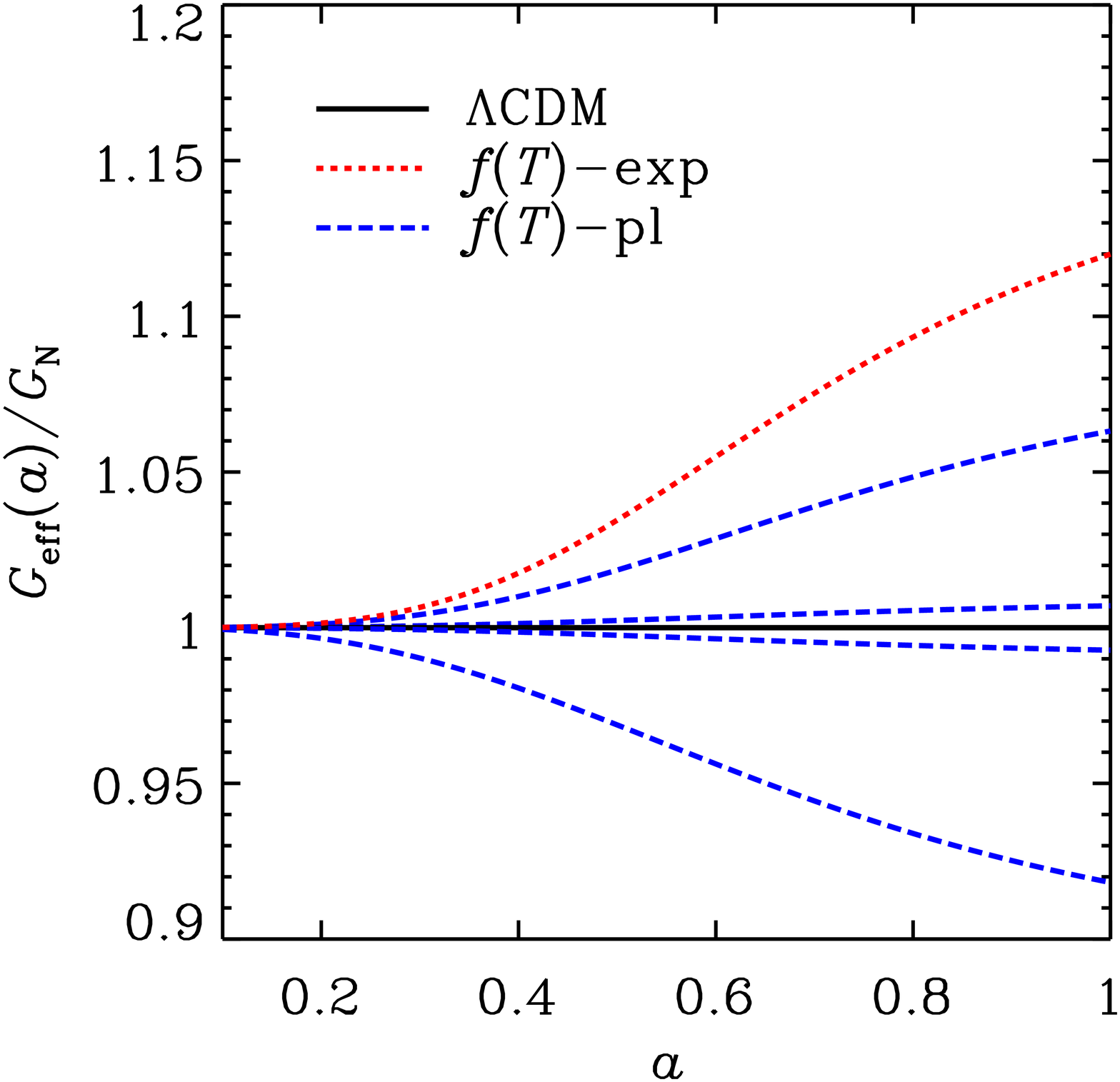}
\caption{Rescaled effective gravitational constant vs the scale factor for the fiducial $f(T)$-exp model (short-dashed, red) and in $f(T)$-pl models (long-dashed, red curves) with $n_T=-0.1$, $-0.01$, $0.01$ and $0.1$ from top to bottom.}\label{fig:Geff}
\end{figure}

It is worth noting, however, that Eq.~\eqref{eq:ftgeff} only holds for $k>k_\mathrm{min}$. On larger scales, the full set of perturbed Einstein's equations has to be solved. For this reason, we have chosen to focus our attention to the sub-horizon limit, thus simplifying the analysis without any loss of the survey constraining power. Indeed, future galaxy survey will not typically be able to probe such extremely large scales, for which alternative techniques are more effective \citep[e.g.][]{Chang:2007xk,Masui:2009cj,Hall:2012wd,Camera:2013kpa}. To give a flavour of the alterations that the $f(T)$ models we analyse bring to the Newtonian constant, we show in Fig.~\ref{fig:Geff} the quantity $\mathcal G_\mathrm{eff}/G_N$ as a function of the scale factor $a$. The solid, black line is the constant value of GR, whilst the exp and pl $f(T)$ models are depicted by the short-dashed, blue and long-dashed, red curves, respectively. The former is calculated with the fiducial values found in Ref.~\citep{Cardone:2012xq}, whilst for the latter we present a few values of $n_T$, specifically $-0.1$, $-0.01$, $0.01$ and $0.1$ from top to bottom. We remind the reader that $n_T=0$ recovers GR.

In order to go from $P^\delta(k,z)$ to $P_\mathrm{obs}(k,\mu;z)$, one has to include anisotropies due to RSDs and account for the fact that the actual measurement concerns the power spectrum of galaxies rather than that of underlying matter fluctuations. Moreover, since the conversion from redshifts to distances is only possible by assuming a reference cosmological model---which can be different from the actual (unknown) one---, a further distortion, referred to as the Alcock-Paczynski effect \citep{Alcock:1979mp}, takes place. The final observed power spectrum then reads \cite{Seo:2003pu,Wang:2006qt,Majerotto:2012mf,Wang:2010gq,Camera:2012sf}
\begin{multline}
P_\mathrm{obs}(k,\mu;z)=\left[\frac{H_\mathrm{ref}(z)}{H(z)}\right]\left[\frac{d_A^\mathrm{ref}(z)}{d_A(z)}\right]^2\\\times[b_g^2(z)+2\mu^2b_g(z)g(z)+\mu^4g^2(z)]\\\times\exp{\left\{-\left[\frac{q\nu c\sigma_z^{\mathrm{sp}}}{H_\mathrm{ref}(z)}\right]^2\right\}}P^\delta(q,\nu;z).\label{eq:pobs}
\end{multline}
Here,
\begin{align}
q(k,\mu,z)&=\mathcal D^{1/2}(\mu,z)k,\\
\nu(k,\mu,z)&=\mathcal D^{-1/2}(\mu,z)[H(z)/H_\mathrm{ref}(z)]\mu\label{eq:defqnu}
\end{align}
and
\begin{equation}
\mathcal D=\left[\frac{d_A(z)}{d_A^\mathrm{ref}(z)}\right]^2+\left\{\left[\frac{H_\mathrm{ref}(z)}{H(z)}\right]^2-\left[\frac{d_A(z)}{d_A^\mathrm{ref}(z)}\right]^2\right\}\mu^2.\label{eq:defdap}
\end{equation}
It is worth a brief comment upon the different terms entering Eq.~\eqref{eq:pobs}. First, the power spectrum is not evaluated directly in $(k,\,\mu)$, rather than in the shifted variables $(q,\,\nu)$ as a consequence of the Alcock-Paczynski effect. Indeed, when the reference cosmology used to measure the power spectrum from the data matches the true one, $H_\mathrm{ref}=H$ and $d_{A}^\mathrm{ref}=d_A$ so that $(q,\,\nu)=(k,\,\mu)$, and the multiplicative bias disappears too. Secondly, the term in the second line is due to RSDs which have been modelled here to linear order. Here, $b_g(z)$ is the galaxy bias, which takes the difference between the galaxy distribution and matter density fluctuations into account. As a matter of fact, more sophisticated expressions could be used to improve the agreement with numerical simulations. However, all of them are very well approximated in the linear r\'egime by our formula. Lastly, the third exponential term accounts for errors in the spectroscopic redshift measurement, 
parameterised here as $\sigma_z^{\mathrm{sp}}$.

\subsection{2D Galaxy Clustering}\label{ssec:clustering2d}
The study of three-dimensional galaxy clustering presented in Sect.~\ref{ssec:clustering3d} has got as basic assumption that we can average the matter power spectrum within each redshift bin. However, the measured redshift is used for both estimating distances, through the radial comoving distance $\chi(z)$, and time, since $z(t)=1/a(t)-1$. In practice, in the $i$th redshift slice, we reconstruct the galaxy power spectrum $P^g(k,z_i)$ by computing correlations amongst galaxy number density fluctuations whose physical separation estimates are functions of the galaxy redshifts. We then relate the reconstructed $P^g(k,z_i)$ to the redshift $z_i$ (usually the centre of the bin). Nonetheless, the sources contained in the volume $V_\mathrm{survey}(z_i)$ have emitted their photons at different instants in the time interval $\Delta t$ centred in $t_i=t(z_i)$. In homogenising everything to the central redshift value, $z_i$, we therefore disregard the time evolution of the underlying matter density field $\delta=\delta\rho_m/\rho_m$. This approximation is harmless provided the width of the redshift slice thin enough so that evolution within the bin is negligible---that is to say, the growth rate is substantially constant. As we will see in Sect.~\ref{sec:results}, this is indeed the case of a spectroscopic galaxy survey, for the spectro-$z$ error, $\sigma_z^{\mathrm{sp}}$, is small, and we can safely consider small-size, sharp-edged redshift slices.

However, in some situations there is no radial information available---or it is poor, consequently meaning that the redshift slices are broad. For instance, this is the case of photometry, where the scatter between the measured and the actual redshift may be large. In this case, we instead deal with projected quantities. Thus, the angular power spectrum $C^g(\ell)$ of galaxy number density fluctuations reads
\begin{equation}
C^g_\ell=4\pi\int\frac{\de k}{k}\left[\mathcal W^g(\ell,k)\right]^2P^g(k,z=0),
\end{equation}
with $\ell$ the angular wavenumber and $\mathcal W^g(\ell,k)$ a proper line-of-site weight function. A widely used simplification is given by so-called Limber's approximation \citep{1953ApJ...117..134L,Kaiser:1987qv}, where $\ell=k\chi$. Limber's approximation is valid when $\ell\gg1$, but it has been shown that the convergence is already good for $\ell\gtrsim10$ \citep[e.g.][]{Hu:2000ee}. Therefore, it is a suitable approximation, since for larger angular scales the cosmic variance uncertainty is dominant. In this limit, and if we can further sub-divide the source sample into some redshift bins, we then have
\begin{equation}
\mathbf C^g_{ij}(\ell)=\int\!\!\de\chi\,\frac{W^g_i(\chi)W^g_j(\chi)}{\chi^2}P^\delta\!\left(\frac{\ell}{\chi},\chi\right),
\end{equation}
with $W^g(\chi)$ defined by
\begin{equation}
W^g[\chi(z)]=H(z)b_g(z)\frac{\de N^{(\mathrm{ph})}}{\de z}(z),
\end{equation}
`ph' denoting photometry. This is usually referred to as redshift tomography, and the two-dimensional galaxy power spectrum is rather a tomographic matrix $\mathbf C^g_{ij}(\ell)$, whose entries are the angular power spectra of each bin.

Lastly, there is a further subtlety that has to be taken into account when dealing with Limber's approximation. Indeed, since it links the angular scale $\ell$ to the physical wavenumber $k$ through the radial comoving distance $\chi$, it is no longer possible to neatly separate linear to non-linear scales as small or large multipoles---conversely to what one does with the three-dimensional $P^\delta(k,z)$. Therefore, we decide to proceed as follows. We now include the non-linear evolution of the matter power spectrum; to do so, we use \textsc{halofit} fitting formul\ae. That the non-linear evolution of density fluctuations in $f(T)$ cosmology follows that of the \lcdm\ model might be seen as a rather strong assumption. However, we believe it acceptable for two reasons: $i)$ Li, Sotiriou \& Barrow~\citep{Li:2011wu} have clearly demonstrated that viable $f(T)$ cosmologies differ from \lcdm\ in the largest, linear scales, otherwise recovering the GR prediction when approaching the non-linear r\'egime; and $ii)$ we anyway limit our analysis to a range of $\ell$'s whereby only mildly non-linear $k$'s are involved, as will be clear in the discussion of the results.

For a square patch of the sky, the Fourier transform leads to uncorrelated modes, provided the modes are separated by $2\pi/\Theta_\mathrm{rad}$, where $\Theta_\mathrm{rad}$ is the side of the square in radians. Then, the Fisher matrix is simply the sum of the Fisher matrices of each $\ell$ mode \citep{Hu:1999ek}, namely
\begin{equation}
\mathbf F_{\alpha\beta}^{{g_\mathrm{2D}}}=f_\mathrm{sky}\sum_{\ell=\ell_\mathrm{min}}^{\ell_\mathrm{max}}\frac{2\ell+1}{2}\mathrm{Tr}\left[\frac{\partial\mathbf C^g(\ell)}{\partial\vartheta_\alpha}\widetilde{\mathbf C^g_\ell}^{-1}\frac{\partial\mathbf C^g(\ell)}{\partial\vartheta_\beta}\widetilde{\mathbf C^g_\ell}^{-1}\right],\label{eq:Fisher}
\end{equation}
where $f_\mathrm{sky}$ is the fraction of the sky covered by the survey under analysis and 
\begin{equation}
\left[\widetilde{\mathbf C^g_\ell}\right]_{ij}=\mathbf C^g_{ij}(\ell)+\frac{1}{N^{(i)}_g}\delta^K_{ij},\label{eq:fisher-clustering2d}
\end{equation}
is the observed (signal plus noise) galaxy angular power spectrum, with $N^{(i)}_g$ the galaxy number density per square arcminute in the $i$th bin and $\delta^K$ the Kronecker delta symbol.

\subsection{Cosmic Shear}\label{ssec:shear}
The presence of intervening matter along the path of photons emitted by distant sources causes gravitational lensing distortions of the high-redshift source images. The weak lensing r\'egime occurs when lensing effects can be evaluated on the null-geodesic of the unperturbed (unlensed) photon \citep{Bartelmann:1999yn}. Such distortions---directly related to the distribution of matter on large scales and to the Universe's geometry and dynamics---can be decomposed into a convergence, $\kappa$, and a (complex) shear, $\gamma=\gamma_1+i\gamma_2$ \citep{Kaiser:1996tp,Bartelmann:1999yn}. Let us now consider a perturbed metric about the flat FLRW background in the longitudinal gauge, viz.
\begin{align}
e^0&=\left(1+2\Phi\right)dt,\\
e^i&=a(t)\left(1+2\Psi\right)dx^i,
\end{align}
where $\Phi$ and $\Psi$ are the two metric potential. For them, $\Phi=-\Psi$ holds in GR and in the absence of anisotropic stress; but this is not, in general, true in extended/modified theories of gravitation. In the sub-horizon r\'egime, we know that matter density fluctuations $\delta$ obey the approximate evolution equation \citep{Zheng:2010am,Li:2011wu,Wu:2012hs}
\begin{equation}
\ddot\delta+2H\dot\delta-4\pi\mathcal  G_\mathrm{eff}\rho_m\delta\simeq0,\label{eq:delta}
\end{equation}
where $\mathcal G_\mathrm{eff}$ is given in Eq.~\eqref{eq:ftgeff}.
To confront our model with weak lensing observations, we have to define the so-called deflecting potential \citep[e.g.][]{Camera:2011ms,Schimd:2004nq,Tsujikawa:2008in,Camera:2011mg}
\begin{equation}
\Upsilon=\frac{\Phi-\Psi}{2},
\end{equation}
and use its Poisson-like equation
\begin{equation}
\nabla^2\Upsilon=4\pi\mathcal G_\mathrm{eff}a^2\rho_m\delta.
\end{equation}
It relates matter density fluctuations to the combination of metric potentials that are responsible for weak gravitational lensing effects. Thence, we can similarly link the power spectrum of the weak lensing source field (namely, the deflecting potential wells) to the three-dimensional matter power spectrum through
\begin{equation}
P^\Upsilon(k,z)=\left[-\frac{3}{2}\ho^2\om(1+z)k^{-2}\frac{\mathcal G_\mathrm{eff}(z)}{G_N}\right]^2P^\delta(k,z).\label{eq:P_Upsilon}
\end{equation}

In the flat-sky approximation, the shear is expanded in its Fourier modes and the two-dimensional angular power spectrum $C^\gamma(\ell)$ is thus given by
\begin{equation}
\langle\gamma(\boldsymbol\ell)\gamma^\ast(\boldsymbol\ell')\rangle={(2\pi)}^2\delta_D(\boldsymbol\ell-\boldsymbol\ell')C^\gamma(\ell),
\end{equation}
with $\ell=|\boldsymbol\ell|$ the angular wavenumber and $\delta_D$ the Dirac delta function. In the case where one has distance information for individual sources, we can use this information for statistical studies. A natural course of action is to divide the survey into slices at different distances, and perform a study of the shear pattern on each slice \citep{Hu:1999ek}. This procedure is the same redshift tomography introduced in Sec.~\ref{ssec:clustering2d}. By doing so, we can construct the tomographic shear matrix $\mathbf C^\gamma(\ell)$, whose elements read
\begin{equation}
\mathbf C^\gamma_{ab}(\ell)=\int\!\!\de\chi\,\frac{W^\gamma_a(\chi)W^\gamma_b(\chi)}{\chi^2}P^\delta\!\left(\frac{\ell}{\chi},\chi\right);\label{eq:tomography}
\end{equation}
from Eq.~\eqref{eq:P_Upsilon}, we have the weak lensing selection (or weight) function in the $a$th redshift bin
\begin{equation}
W_a^\gamma(\chi)=\frac{3}{2}\ho^2\om\frac{\chi}{a(\chi)}\frac{\mathcal G_\mathrm{eff}(\chi)}{G_N}\int_\chi^\infty\de\chi'\,\frac{\chi'-\chi}{\chi'}\frac{\de N^\mathrm{(ph)}_a}{\de\chi'}\label{eq:W(z)}
\end{equation}
with
\begin{equation}
\frac{\de N^\mathrm{(ph)}_a}{\de\chi}=\frac{\de N^\mathrm{(ph)}_a}{\de z}\frac{\de z}{\de\chi}
\end{equation}
the redshift distribution of the sources. Here, as in the case of two-dimensional galaxy clustering, $\de N_i/\de\chi$ is basically the probability of finding a source within the $a$th bin, and, as such, it must have unity area. Also, we have again used Limber's approximation.

For cosmic shear tomography, the Fisher matrix $\mathbf F^\gamma$ is functionally identical to that of two-dimensional angular clustering in Eq.~\eqref{eq:fisher-clustering2d},
where, now, the observed (signal plus noise) shear angular power spectrum reads
\begin{equation}
\left[\widetilde{\mathbf C^\gamma_\ell}\right]_{ab}=\mathbf C^\gamma_{ab}(\ell)+\frac{{\sigma_\gamma}^2}{N_a}\delta^K_{ab},
\end{equation}
with $\sigma_\gamma\simeq0.3$ the galaxy-intrinsic shear rms in one component.

\subsection{Galaxy-Shear Cross-Correlation}\label{ssec:cross}
Thanks to the formalism described in Sect.~\ref{ssec:clustering2d}, two-dimensional galaxy clustering also enable us to estimate its cross-correlation with the cosmic shear signal. It can be easily computed through
\begin{equation}
\mathbf C^{g\gamma}_{ia}(\ell)=\int\!\!\de\chi\,\frac{W^g_i(\chi)W^\gamma_a(\chi)}{\chi^2}P^\delta\!\left(\frac{\ell}{\chi},\chi\right).
\end{equation}
Then, the observed (signal plus noise) cross-correlation is
\begin{equation}
\left[\widetilde{\mathbf C^{g\gamma}_\ell}\right]_{ia}=\mathbf C^{g\gamma}_{ia}(\ell),
\end{equation}
since clustering and shear noise contributions do not correlate.

\section{Results and Discussion}\label{sec:results}
First of all, we need to specify a reference survey whose constraining power we want to test with the Fisher matrix formalism sketched in Sect.~\ref{sec:observables}. For our purpose, we find that a \textit{Euclid}-like experiment \citep{EditorialTeam:2011mu,Amendola:2012ys} perfectly suits our endeavour, since it will perform both (spectroscopic) galaxy-clustering and (photometric) cosmic-shear measurements. \textit{Euclid} is an ESA medium class space mission selected in October $2011$ in the Cosmic Vision $2015$-$2025$ programme, and it results of the merging of the DUNE and SPACE missions. The \textit{Euclid} mission aims at understanding why the expansion of the Universe is accelerating and what is the nature of the source responsible for this acceleration. Therefore, it is in thorough agreement with the effort of our work.

\begin{figure}
\centering
\includegraphics[width=0.5\textwidth]{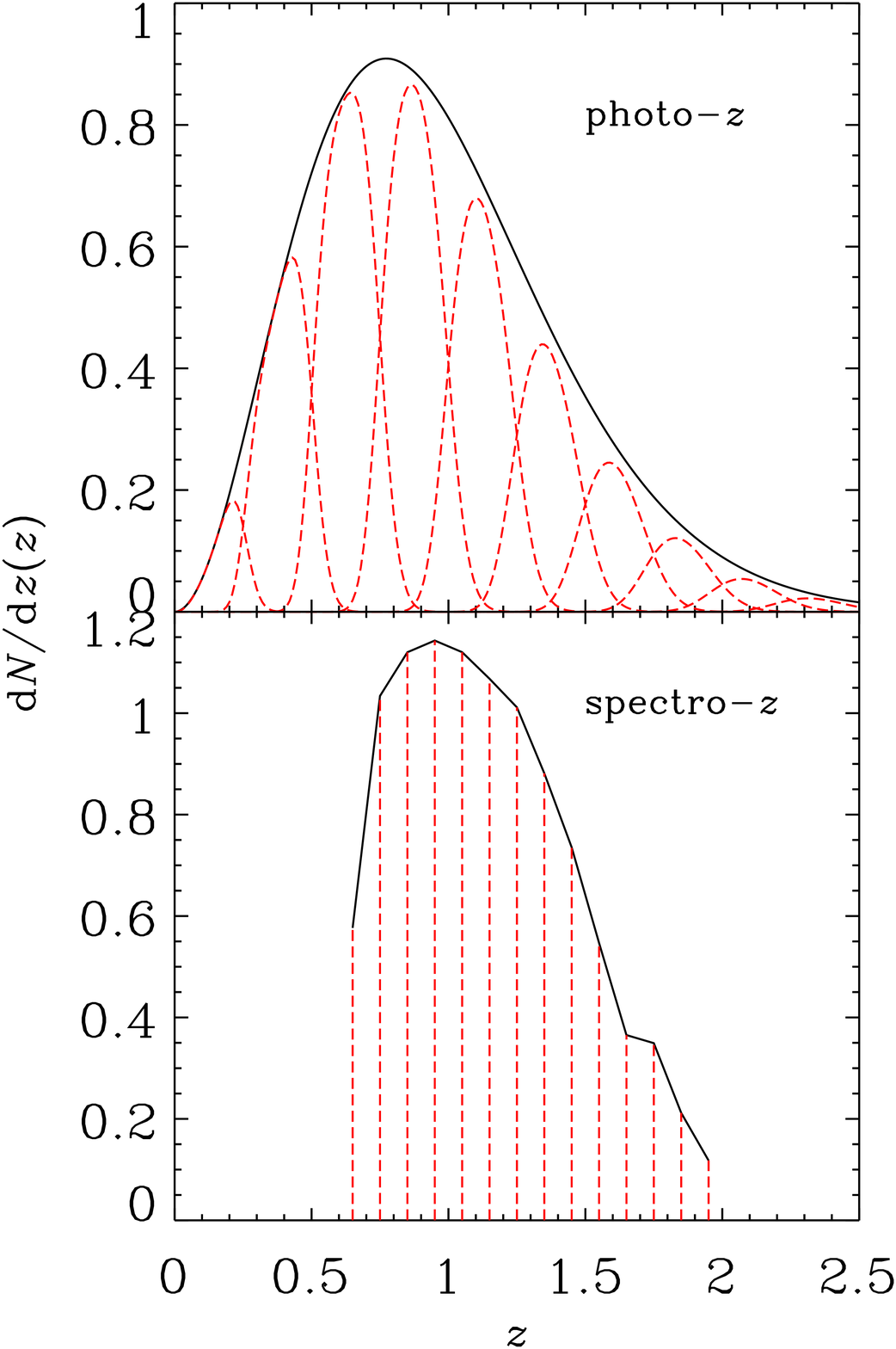}
\caption{Normalised source redshift distributions $\de N/\de z(z)$. \textit{Bottom panel:} spectroscopic galaxy survey (solid, black) and its fourteen bins (dashed, red). \textit{Top panel:} photometric imaging survey (solid, black) and its ten bins (dashed, red).}\label{fig:dNdz}
\end{figure}

The spectroscopic survey (hereafter \textit{Euclid}-sp) will measure galaxy redshifts in the infrared band $0.9-2\,\mu\mathrm m$ for $\sim 65$ million galaxies using a slitless spectrograph relying on the detection of emission lines in the galaxy spectra. In the chosen wavelength range, the most favourable line will be the H$\alpha$ line redshifted to $0.7\le z\le2$. We divide this range in equally spaced bins of width $0.1$. This is much larger than the typical errors on spectro-$z$'s, which we take as $\sigma_z^{\mathrm{sp}}=0.001(1+z)$. By doing so, redshift errors are much smaller than the bin width, the constraints on cosmological parameters from different bins are thus independent from each other. Therefore, we can first marginalise over the bias in each bin and then sum the resulting Fisher matrices to get the final $\mathbf F^{g_\mathrm{3D}}$. The redshift distribution should be computed taking into account the instrumental set-up and its efficiency coupled to a model of the number density of H$\alpha$ emitters. Following Ref.~\cite{Majerotto:2012mf,Wang:2010gq,Camera:2012sf} and the definition study report \citep{EditorialTeam:2011mu}---which we refer to for details---we use the distribution of H$\alpha$ emitters of Ref.~\citep{2010MNRAS.402.1330G} and weight it according to a proper flag \citep{2010PASP..122..827G}. This eventually provides the $\de N^\mathrm{(sp)}/\de z(z)$ profile shown in the bottom panel of Figure~\ref{fig:dNdz}.

For what concerns the photometric measurements (\textit{Euclid}-ph), we compute our results for a $15,000\,\mathrm{deg}^2$ cosmic-shear experiment. The source distribution over redshifts has the form \citep{1994MNRAS.270..245S}
\begin{equation}
\frac{\de N^\mathrm{(ph)}}{\de z}(z)\propto z^2e^{-\left(\frac{z}{z_0}\right)^{1.5}},\label{eq:n_z-Euclid}
\end{equation}
where $z_0=z_m/1.41$, and $z_m=0.9$ is the median redshift of the survey. The number density of the sources, with estimated photometric redshift and shape, is $30$ per square arcminute. To perform the tomographic analyses outlined in Sects~\ref{ssec:clustering2d} to \ref{ssec:cross}, we divide the redshift distribution of sources into ten redshift bins. However, the Euclid imaging survey will only provide photometric-redshift measurements, which are known to be less accurate than those obtained from spectroscopy. The scatter between the true redshift and the photometric estimate is assumed to be of order $3\%$ and scale linearly with $z$, that is to say $\sigma_z^\mathrm{ph}=0.03(1+z)$. The top panel of Fig.~\ref{fig:dNdz} illustrates the total $\de N^{(ph)}_g/\de z$ (solid, black) and the ten photometric-redshift bins we use (dashed, red).

\subsection{3D Galaxy Clustering Constraints}\label{ssec:res-clustering3d}
Let us start by examining the constraints on $f(T)$ gravity from the three-dimensional galaxy power spectrum alone. First, we consider the exp model and estimate the Fisher matrix with respect to the parameters $\boldsymbol\vartheta=\{\om,\,h,\,n_T,\,p_T,\,n_s,\,\sigma_8\}$, and we marginalise over the bias $b_g(z_i)$ in each redshift bin. As fiducial values, we choose $\{\om,\,h,\,n_T,\,p_T\}=\{0.287,\,0.731,\,0.736,\,-0.100\}$, according to the results in Ref.~\citep{Cardone:2012xq}, whilst we set $\{n_s,\,\sigma_8\}=\{0.820,\,0.9608\}$, in agreement with the WMAP9 constraints. As done in Ref.~\citep{Majerotto:2012mf}, the fiducial bias values for each bin have been set following Ref.~\citep{2010MNRAS.405.1006O}.

Despite background parameters $\om$ and $h$ and the power spectrum related quantities $n_s$ and $\sigma_8$ are well constrained, confidence ranges for $f(T)$ parameters are quite broad. In particular, we find $\sigma(n_T)=3.0$ and $\sigma(p_T)=3.7$. Such a result can be qualitatively explained as follows. Over the redshift range $0.7-2.0$, the term $T_0/T=E^{-2}(z)$ quickly decreases so that the exponential in Eq.~\eqref{eq:ftexp} approaches unity. Hence, the $f(T)$ term in the Lagrangian becomes subdominant. Such a behaviour holds whatever are the values of $n_T$ and $p_T$, thus explaining why it is so difficult to constrain these parameters using three-dimensional clustering alone. However, it is worth emphasising, that this result is mainly due to the redshift range investigated rather than the observational probe adopted---as will be clear in the following section. One could na\"ively expect that shifting the median survey redshift to a smaller $z$ would improve the constraining power, since the power spectrum would be more sensitive to the $f(T)$ parameters. However, to have a lower median redshift for the survey, one should change the instrumental set-up and rely on different emission lines, that is to say different kinds of target galaxies. As a consequence, the redshift distribution would also change, and it is not possible \textit{a priori} to infer whether the constraints will improve or degrade.
\begin{table}
\caption{\label{tab:constraints-clustering3d}Forecast $1\sigma$ marginal errors on $f(T)$ model parameters from three-dimensional galaxy clustering alone.}
\begin{ruledtabular}
\begin{tabular}{lccc}
&\multicolumn{2}{c}{exp}&\multicolumn{1}{c}{pl}\\
\cline{2-3}
& $n_T$ & $p_T$ & $n_T$ \\
\hline
$g_\mathrm{3D}$ \textit{Euclid}-sp & 3.0 & 3.7 & 0.021
\end{tabular}
\end{ruledtabular}
\end{table}

Fisher matrix forecasts depend not only on the observational probe adopted and the precision in the measurements, but also on the fiducial cosmological model. A interesting example is provided here by the results for the $f(T)$-pl model. We assume $n_T=0$, which \textit{de facto} implies a \lcdm\ scenario. As expected, the constraints on the standard parameters $\{\om,\,h,\,n_s,\,\sigma_8\}$ are comparable with those obtained for the exp model and other in the literature. Moreover, the slope $n_T$ of the $f(T)$ term is now well constrained, with $\sigma(n_T)=0.021$. This encouraging result suggests that three-dimensional galaxy clustering alone is able to detect torsion gravity departures from the GR based \lcdm\ scenario.

\subsection{2D Galaxy Clustering Constraints}\label{ssec:res-clustering2d}
Now, we analyse the results from two-dimensional (photometric) galaxy clustering. Since the redshift slices are broader than before, we can no more marginalise over the bias amplitude in each bin, and then sum over the bins. Hence, in this case the parameter set, for example for the $f(T)$-exp model, is $\boldsymbol\vartheta=\{\om,\,h,\,n_T,\,p_T,\,n_s,\,\sigma_8,\,\mathbf b_g\}$; $\mathbf b_g$ is a vector of nuisance parameters which account for the bias amplitude in each redshift bin. This is slightly different from what done in the 3D case, where a nuisance bias parameter is included in each redshift binned Fisher matrix, then marginalised over to eventually sum all the marginalised Fisher matrices. This happens because in the 3D case one considers the various redshift bins as uncorrelated volumes of the Universe, whereas in the 2D case one in principle also includes cross-correlations between bins. Anyway, we emphasise that this is somehow an over-conservative approach, because, even though we do not exactly know the H$\alpha$ galaxy bias, it cannot freely vary in each bin. Nevertheless, we decide to proceed so also to safely deal with our ignorance of the halo bias in $f(T)$ gravity. According to Ref.~\citep{EditorialTeam:2011mu}, the angular multipoles that will be probed by \textit{Euclid} are in the range $\ell\in[5,\,5000]$. However, we find this assumption rather too optimistic for the present case: on the one hand, for $\ell\lesssim10$, Limber's approximation is less safe \citep{Kaiser:1996tp,LoVerde:2008re}; on the other hand, at very small angular scales (large $\ell$'s), non-linear effects---as well as feedback from baryonic physics---became non negligible \citep{White:2004kv,Zentner:2012mv}. Therefore, we decide to scrutinise three different scenarios, dubbed \textit{Euclid}-ph I, II and III, where $\ell\in[10,\,1000]$, $[10,\,3000]$ and $[5,\,5000]$, respectively.

In Table~\ref{tab:constraints-clustering2d}, we present the forecast $68.3\%$ marginal errors on $f(T)$-exp and $f(T)$-pl model parameters. It is straightforward to notice that, as expected, the wider the range of angular multipoles, the tighter the constraints. Besides, it is interesting to verify the explanation presented in Sect.~\ref{ssec:res-clustering3d} on the reason for why the $f(T)$-exp parameters were poorly constrained by three-dimensional galaxy clustering. Indeed, the range of redshifts probed by the \textit{Euclid} imaging survey is wider than the $0.7\leq z\leq2.0$ interval motivated by H$\alpha$ line spectroscopy. As a consequence, forecast marginal errors obtained with now are $3.5$ to $>8$ times more stringent than those got with $\mathbf F^{g_\mathrm{3D}}$, for $n_T$, and $3$ to $6$ times for $p_T$.
\begin{table}
\caption{\label{tab:constraints-clustering2d}Forecast $1\sigma$ marginal errors on $f(T)$ model parameters from two-dimensional galaxy clustering alone.}
\begin{ruledtabular}
\begin{tabular}{lccc}
&\multicolumn{2}{c}{exp}&\multicolumn{1}{c}{pl}\\
\cline{2-3}
& $n_T$ & $p_T$ & $n_T$ \\
\hline
$g_\mathrm{2D}$ \textit{Euclid}-ph I & 0.86 & 1.2 & 0.12 \\
$g_\mathrm{2D}$ \textit{Euclid}-ph II & 0.54 & 0.84 & 0.050 \\
$g_\mathrm{2D}$ \textit{Euclid}-ph III & 0.37 & 0.61 & 0.035
\end{tabular}
\end{ruledtabular}
\end{table}

Regarding the $f(T)$-pl model, constraints from two-dimensional angular power spectrum in the most conservative \textit{Euclid}-ph I configuration are almost one order of magnitude weaker than the three-dimensional case. A reason for this can be understood by looking at the correlations amongst $n_T$ and the other parameters, namely the $r(n_T,\vartheta_\alpha)$ coefficients. They read $0.78$, $-0.835$, $0.73$ and $-0.14$, for $\om$, $h$, $n_s$ and $\sigma_8$, respectively. This means that the slope of the power law modification to the teleparallel gravity Lagrangian is degenerate with almost all the standard cosmological parameters---particularly those related to the background expansion history. This happens because the functional form of the $f(T)$-pl model is basically a rescaled version of the Hubble parameter. (Please remind that $T=-6H^2$.) Thus, the non-standard parameter $n_T$ simply alters the evolution in redshift of the torsion scalar, $T$, without introducing any peculiar behaviour, as is instead the case of the $f(T)$-exp model. Nonetheless, things are better if we increase the analysed $\ell$ range. For example, already with the \textit{Euclid}-ph II configuration, we have a promising $\sigma(n_T)=0.050$, and indeed  $r(n_T,\vartheta_\alpha)$ coefficients are now $0.59$, $-0.53$, $0.46$ and $0.34$.

\subsection{Cosmic Shear Constraints}\label{ssec:res-shear}
Let us now move to analyse $f(T)$ model parameter constraints coming from cosmic shear alone. In Table~\ref{tab:constraints-shear}, we present the forecast $68.3\%$ marginal errors on cosmological parameters for both $f(T)$-exp and  $f(T)$-pl models. Again, the wider the range of angular multipoles, the tighter the constraints. Besides, we can easily see that they are overall better than in the case of two-dimensional clustering. This behaviour has a straightforward reason. Indeed, the weak-lensing weight function of Eq.~\eqref{eq:W(z)} does have a further (and more direct) dependence upon $\mathcal G_\mathrm{eff}$, compared to galaxy clustering. Thus, cosmic shear---and weak lensing effects more generically---is more effective in detecting modified gravity effects.
\begin{table}
\caption{\label{tab:constraints-shear}Forecast $1\sigma$ marginal errors on $f(T)$ model parameters from cosmic shear tomography alone.}
\begin{ruledtabular}
\begin{tabular}{lccc}
&\multicolumn{2}{c}{exp}&\multicolumn{1}{c}{pl}\\
\cline{2-3}
& $n_T$ & $p_T$ & $n_T$ \\
\hline
$\gamma$ \textit{Euclid}-ph I & 0.62 & 0.90 & 0.11 \\
$\gamma$ \textit{Euclid}-ph II & 0.40 & 0.59 & 0.096 \\
$\gamma$ \textit{Euclid}-ph III & 0.31 & 0.48 & 0.088
\end{tabular}
\end{ruledtabular}
\end{table}

\subsection{Combined Constraints}\label{ssec:res-tot}
After having analysed the constraining power of galaxy clustering and cosmic shear singularly, having thus understood the most important aspects and peculiarities of the two probes, it is now time to look at the combination of the two. To better investigate the effect of modified torsion gravity on weak lensing, in Sect.~\ref{ssec:res-shear} we have presented the results for three different \textit{Euclid}-like scenario. However, we now restrict ourselves to the most conservative case. Indeed, with $\ell_\mathrm{min}=10$ we are confident that Limber's approximation is used in its r\'egime of validity. Moreover, we do not want our results to rely on non-linear scales, whose dynamics and growth of perturbations has not yet been studied in $f(T)$ cosmology. Thus, $\ell_\mathrm{max}=1000$ better suits our purpose.

The only consistent way to combine clustering and shear forecasts, as discussed in Ref.~\citep{Giannantonio:2011ya}, is by using angular power spectra for both. By doing so, we construct a new Fisher matrix containing not only all the $\mathbf C^g$ and $\mathbf C^\gamma$ spectra, but also their cross-correlations. Hence, we can build a combined tomographic matrix
\begin{equation}
\widetilde{\mathbf C_\ell}=\left(
\begin{array}{cc}
\widetilde{\mathbf C^g_\ell} & \widetilde{\mathbf C^{g\gamma}_\ell} \\
\widetilde{\mathbf C^{g\gamma}_\ell} & \widetilde{\mathbf C^\gamma_\ell}
\end{array}
\right),
\end{equation}
and its corresponding Fisher matrix takes again the same form as Eq.~\eqref{eq:Fisher} \citep{Hu:2003pt}. By doing so, we obtain the constraints presented in Table~\ref{tab:constraints-combined2d}.
\begin{table}
\caption{\label{tab:constraints-combined2d}Forecast $1\sigma$ marginal errors on $f(T)$ model parameters from the combination of two-dimensional galaxy clustering and cosmic shear.}
\begin{ruledtabular}
\begin{tabular}{lccc}
&\multicolumn{2}{c}{exp}&\multicolumn{1}{c}{pl}\\
\cline{2-3}
& $n_T$ & $p_T$ & $n_T$ \\
\hline
$\gamma+g_\mathrm{2D}$ \textit{Euclid}-ph I & 0.063 & 0.14 & 0.0097
\end{tabular}
\end{ruledtabular}
\end{table}

\begin{figure*}
\centering
\includegraphics[width=0.475\textwidth]{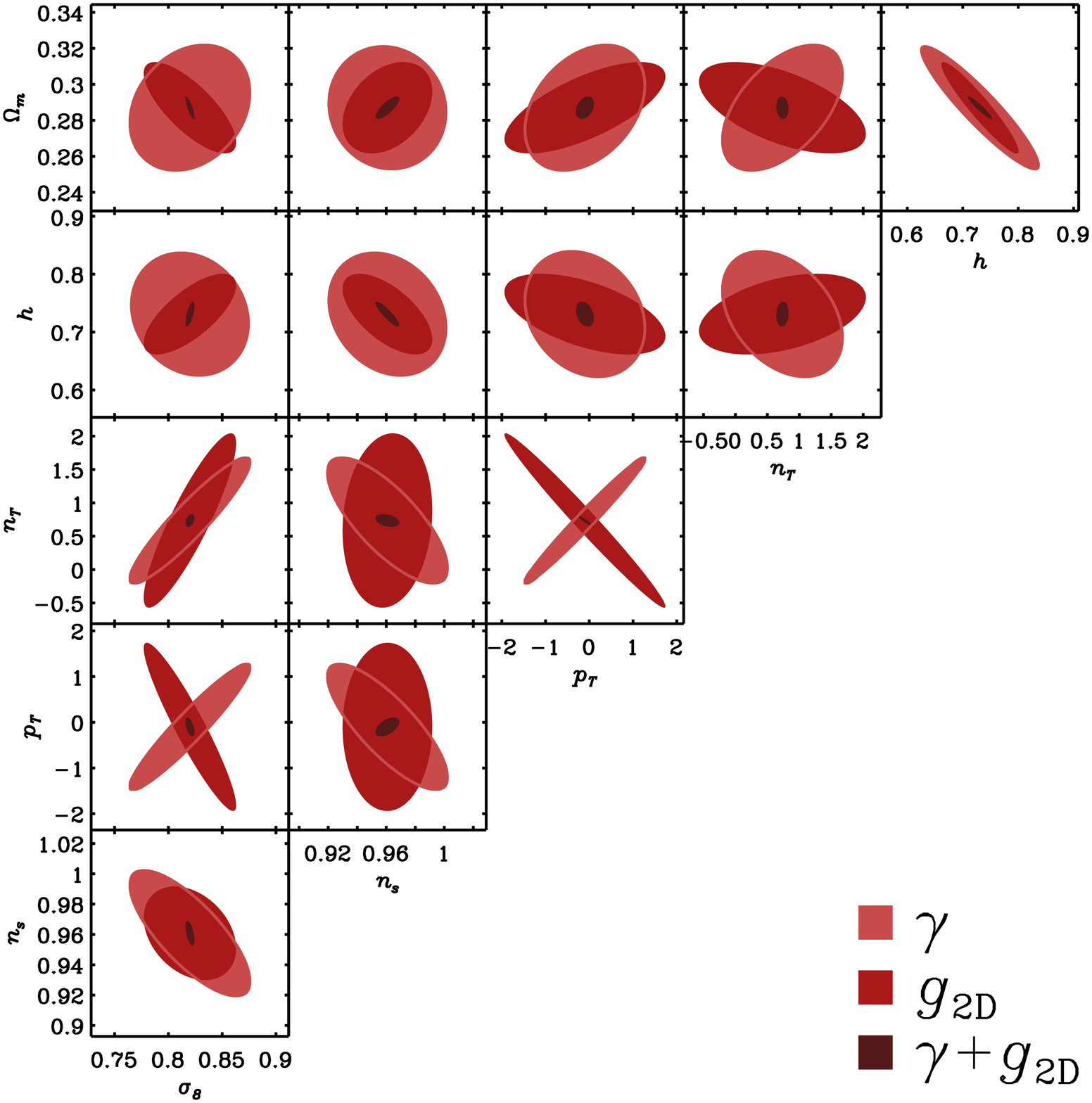}
\includegraphics[width=0.475\textwidth]{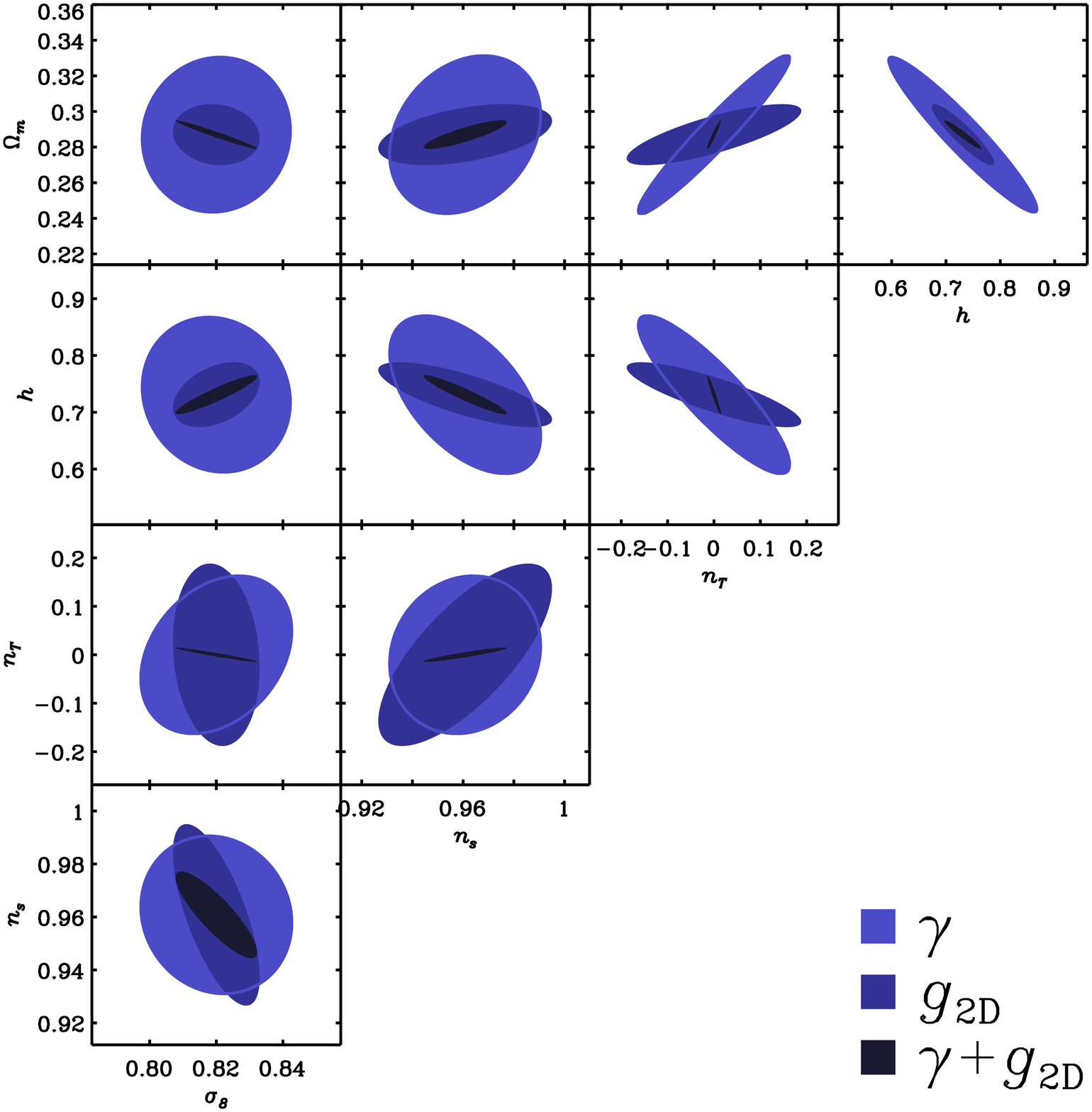}
\caption{Marginal error contours in the two-parameter plane for $f(T)$-exp (left panel) and $f(T)$-pl (right panel) model parameters.}\label{fig:ellipses-tot}%\label{fig:ellipses-pl-tot}
\end{figure*}

It is immediate that the combination of the two probes greatly enhances the constraining potential of the survey. This is due to the fact that parameter degeneracies in $\mathbf F^{g_\mathrm{2D}}$ and $\mathbf F^\gamma$ are almost always `perpendicular'---in the sense that the correlation coefficients of galaxy clustering and cosmic shear analyses have opposite sign. Thence, all parameter errors shrink. In particular, we have that the $1\sigma$ marginal error on $f(T)$-exp model parameters are $\sigma(n_T)=0.063$ and $\sigma(p_T)=0.14$. Even more impressively, for the $f(T)$-pl parameter we obtain $\sigma(n_T)=0.0097$, more than twice more stringent than what obtained with three-dimensional galaxy clustering, and almost $12$ times better compared to photometric probes alone. All this can be more easily seen in Fig.~\ref{fig:ellipses-tot}, which respectively show the forecast $1\sigma$ two-parameter marginal contours on $f(T)$-exp and $f(T)$-pl model parameters in the $(\vartheta_\alpha,\,\vartheta_\beta)$-planes for galaxy clustering (light colours) and cosmic shear (darker colours) alone and combined (smallest and darkest ellipses). The different---often substantially orthogonal---orientations of the error ellipses demonstrate how effective is the combination of galaxy clustering and cosmic shear tomography for our science case. This is a general trend, but it is even more useful for the modified torsion gravity non-standard parameters.

Such a spectacular behaviour is due to the fact that BAOs and RSDs are highly complementary to weak lensing, especially in the presence of uncertainties of photo-$z$ errors and inaccurate knowledge of galaxy clustering bias \citep{Zhan:2006gi,EditorialTeam:2011mu,Amendola:2012ys}. Galaxy clustering data measure the `dark fluid' equation of state at higher redshift than SNeIa, used so far to this task \citep[cfr.][]{Cardone:2012xq}. Besides, clustering is a probe for the evolution of matter fluctuations; and thus, through the Poisson equation, of the Newtonian potential $\Phi$---and of the modifications occurring from modified torsion gravity to the Newtonian gravitational constant, i.e.\ $\mathcal G_\mathrm{eff}(z)$. On the other hand, weak lensing is sensitive, through the deflecting potential $\Upsilon$, to the sum of the two metric potentials, which are equal in GR but not in more general gravity theories. As a consequence, the sensitivity to beyond-GR growth parameters mostly comes from weak lensing, 
which provide the only direct measurements of growth (without biasing) \citep{Weinberg:2012es}. In other words, constraints on modifications to gravity mostly depend on the errors on cosmic shear---except when intrinsic parameter degeneracies wreak havoc the weak lensing constraining potential. Conversely, these constraints are very weakly sensitive to the BAO errors, showing that the uncertainties are dominated by the growth measurements themselves rather than residual uncertainty in the expansion history.

As a final remark, the next generation of large-scale experiments aiming at understanding the nature of present-day cosmic acceleration seek much higher precision than those carried out to date. Therefore, the risk of being limited or biased by systematic errors is much higher. Conclusions about cosmic acceleration will be far more convincing if they are reached independently by methods with different systematic uncertainties. Hence, measuring angular and tracer dependence of the clustering signal and testing redshift scaling of cosmic shear as we do here is not only more effective, but also safer.

\section{Bayesian Model Selection}\label{sec:lnB}
The main purpose of this paper is to study the detectability of modified torsion gravity signatures by exploiting the potential of future surveys probing the large-scale structure of the cosmos. To this aim, we have hitherto analysed to which degree of accuracy the $f(T)$ model parameters can be constrained by the \textit{Euclid} satellite, as a reference survey. Nonetheless, there is in a sense a higher-level question than parameter estimation: model selection. In the case of the $f(T)$-pl model, we can recast the present analysis as a comparison between the concordance \lcdm\ cosmological model, where gravity is described by standard GR and the present-day cosmic acceleration is caused by a cosmological constant term, and a modified teleparallel scenario, whereby the gravity Lagrangian contains a higher order term, $\alpha(-T)^{n_{T}}$, in the torsion scalar, $T$. In terms of model parameters, this latter perspective does include the former, where the slope of the additional term vanishes, namely $n_T=0$.

When performing standard parameter estimation, we assume a theoretical model within which we interpret the data. Conversely, in model selection what we want to know is which theoretical framework is preferred given the data. Clearly enough, if the alternative model had more parameters than the standard one, chi-squared analysis would be of little use, because it would always reduce if we added more parameters (i.e.\ degrees of freedom). Otherwise, Bayesian analysis provides a useful Occam's razor, known as the Bayes factor, $B$. It involves the computation of the Bayesian evidence \citep[often called \textit{marginal likelihood} or \textit{model likelihood}; cfr.][\S~4.2]{Trotta:2008qt}. For a model $\mathcal M$, it is defined as a marginalisation over its $m$ parameters $\boldsymbol\vartheta$, viz.\
\begin{equation}
\mathcal Z(\mathbf d|\mathcal M)=\int\!\!\de^m\vartheta\,\mathcal L(\boldsymbol\vartheta)\pi(\boldsymbol\vartheta|\mathcal M);
\end{equation}
here, $\mathcal L(\boldsymbol\vartheta)\equiv p(\mathbf d|\boldsymbol\vartheta,\mathcal M)$ is the likelihood function of the parameters (which equals the probability for the model parameters given the data $\mathbf d$) and the prior $\pi(\boldsymbol\vartheta|\mathcal M)$ encodes our status of knowledge before seeing the data.

Let us now consider two competing models $\mathcal M_1$ and $\mathcal M_2$, the former nested in the latter. That is to say, $\mathcal M_1$ is simpler, because the set of its parameters $\{\vartheta_{\alpha_1}\}$ is contained into the $\mathcal M_2$ parameter set $\{\vartheta_{\alpha_2}\}$, with $\alpha_1$ running from $1$ to $m_1$, $\alpha_2$ from $1$ to $m_2$, and with $m_2>m_1$ by definition. In such a situation, one can compute the Bayes factor
\begin{equation}
B=\frac{p(\mathcal M_1|\mathbf d)}{p(\mathcal M_2|\mathbf d)},\label{eq:B_def}
\end{equation}
which is the ratio of the two corresponding posterior evidence probabilities. The posterior probability for each model $\mathcal M_i$ is given by Bayes' theorem,
\begin{equation}
p(\mathcal M_i|\mathbf d)=\frac{\mathcal Z(\mathbf d|\mathcal M_i)\pi(\mathcal M_i)}{p(\mathbf d)}.\label{eq:posterior}
\end{equation}

If we have no \textit{a priori} preferences towards one specific model, this will translate into the choise of non-committal priors $\pi(\mathcal M_1)=\pi(\mathcal M_2)=1/2$. Hence, the ratio of the posterior evidence probabilities Eq.~\eqref{eq:posterior} reduces to the ratio of the evidences. Ref.~\citep{Heavens:2007ka} showed that in $i)$ the Laplace approximation, where the expected likelihoods are given by multivariate Gaussians, and $ii)$ if one considers $\langle B\rangle$ as the ratio of the expected values, rather than the expectation value of the ratio, one eventually gets
\begin{equation}
\langle B\rangle=\frac{\sqrt{\det\mathbf F_2}}{\sqrt{\det\mathbf F_1}}\left(2\pi\right)^{-l/2}\prod_{q=1}^l\pi\left(\vartheta_{\alpha_1+q}\right)e^{-\boldsymbol \delta_\vartheta\cdot\mathbf F_2\cdot\boldsymbol \delta_\vartheta/2}.\label{eq:B}
\end{equation}
Here, $\mathbf F_i$ is the Fisher matrix relative to the $i$th model, $l=m_2-m_1$ is the number of extra parameters, and $\boldsymbol \delta_\vartheta$ is the vector of the parameter shifts. These shifts appear because, if the correct underlying model were $\mathcal M_2$, the maximum of the expected likelihood would not, in principle, be at the correct parameter values of $\mathcal M_1$ \citep[see Fig.~1 of][]{Heavens:2007ka}. The $m_1$ parameters of $\mathcal M_1$ shift their values to compensate the fact that $\vartheta_{\alpha_1+1},\ldots,\vartheta_{\alpha_1+l}$ are kept fixed at some incorrect fiducial value---most of times, as is in our $f(T)$-pl case, simply $\vartheta_{\alpha_1+1}=\ldots=\vartheta_{\alpha_1+l}=0$. The shifts can be computed under the assumption of a multivariate Gaussian distribution \citep{Taylor:2006aw}, and read
\begin{equation}
\boldsymbol \delta_\vartheta=-{\mathbf F_1}^{-1}\cdot\mathbf G_2\cdot\boldsymbol \delta_\psi,\label{eq:shifts}
\end{equation}
with $\mathbf G_2$ a subset of the $\mathcal M_2$ Fisher matrix and $\boldsymbol \delta_\psi$ the shifts of the $l$ extra parameters $\boldsymbol\psi$. So-called `Jeffreys' scale' gives empirically calibrated levels of significance for the strength of evidence \citep{Jeffreys:1961}. A recent version of Jeffreys' scale sets $|\ln B|<1$ as `inconclusive' evidence in favour of a model, $1<|\ln B|<2.5$ as `positive', $2.5<|\ln B|<5$ as `moderate', and $|\ln B|>5$ as `strong' \citep{Trotta:2005ar}.

\begin{figure}
\centering
\includegraphics[width=0.5\textwidth]{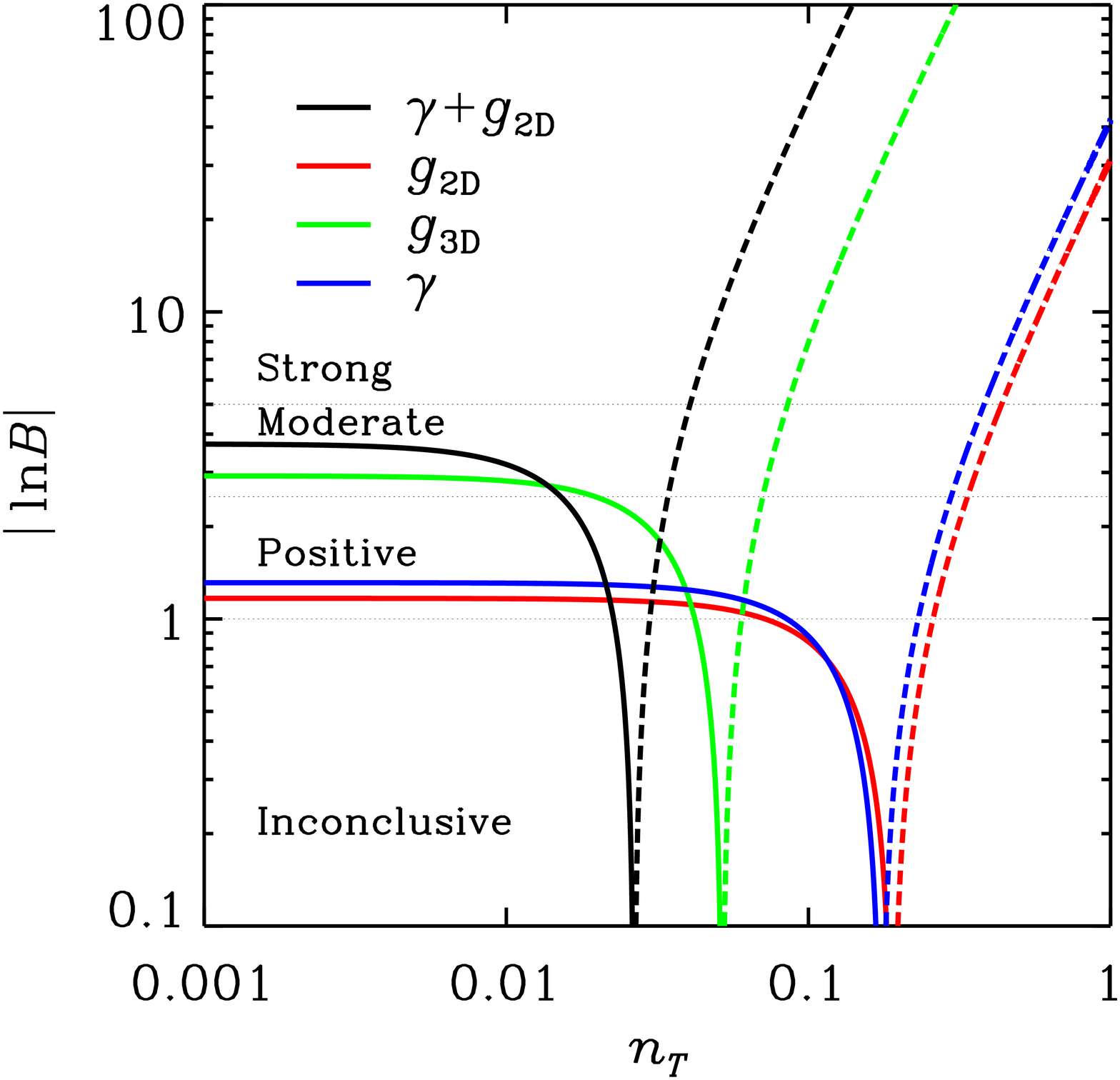}
\caption{$|\ln B|$ versus the $f(T)$-pl $n_T$ parameter for two- and three-dimensional galaxy clustering (red and blue curves, respectively) and cosmic shear alone (red curve), and for the full, combined two-dimensional clustering and shear (black curve). Solid (dashed) curves refer to $B>1$ ($B<1$). Jeffreys' scale confidence levels fall between horizontal thin, dotted lines.}\label{fig:lnB}
\end{figure}

Fig.~\ref{fig:lnB} shows the Bayes factor, $|\ln B|$, as a function of the extra-\lcdm\ parameter $n_T$, for the case of the $f(T)$-pl model. Red and blue lines respectively refer to the use of galaxy clustering and cosmic shear solely, whilst the combination of the two yields the green curve. Horizontal thin, dotted lines indicate the boundaries of Jeffreys' scale confidence levels. When the simpler model, $\mathcal M_1$, is preferred by the data, the ratio in Eq.~\eqref{eq:B_def} is larger than unity, $|\ln B|$ is thus positive (solid lines). Instead, if $\mathcal M_2$, the more complex model, has got a larger posterior evidence probability, Eq.~\eqref{eq:B_def} is $<1$ and the graph in Fig.~\ref{fig:lnB} is consequently negative (dashed lines).

To make a clarifying example, if the reference \textit{Euclid}-like survey were to measure a $n_T$ value of $0.08$, this would imply $\ln B=-4.0$, $0.96$ and $1.0$, for sole three- and two-dimensional clustering and shear, respectively. Occam's razor for cosmic shear would therefore give a \textit{positively} favour \lcdm---that is to say, the \textit{Euclid}-like experiment would not be able to decisively state the goodness of \lcdm\ over modified torsion gravity. Even worse, two-dimensional clustering would be \textit{inconclusive}, because of its weak constraints on $n_T$. Oppositely, $\sigma(n_T)=0.021$ coming from galaxy clustering would provide a \textit{moderate} evidence towards $f(T)$ gravity. However, the complementarity of galaxy clustering and cosmic shear is such that the combined forecast marginal errors are much tighter than what obtained by single probes. Indeed, for a measure of $n_T=0.08$, we would have $\ln B=-30$ (black line, dashed branch), which falls into the \textit{strong} confidence level of Jeffreys' scale---a strong observational evidence in favour of the $f(T)$-pl model. In other words, the odds for modified torsion gravity to \lcdm\ would be $\sim10^{13}:1$.

\section{Conclusions}\label{sec:conclusions}
In this paper, we have analysed cosmological models derived from modified torsion gravity theories, commonly referred to as $f(T)$ cosmologies. Our aim has been to investigate the detectability of $f(T)$ signatures via measurements of the growth and dynamics of the large-scale cosmic structure. The motivation for this work effort is twofold. First, amongst the plethora of modified gravity theories proposed as solutions to the dark energy puzzle, $f(T)$ cosmologies represent an intriguing scenario, for they still give second order equations in field derivatives, oppositely to the fourth order equations of $f(R)$ gravity. Moreover, $f(T)$ models violate Lorentz invariance, and do not therefore belong to the family of Horndeski theories. Thus, they are worth being scrutinised, since constraints on Horndeski Lagrangian will not be able to confirm or rule out modified torsion gravity. Secondly, a vast number of experiments aiming at probing the properties of the Universe's large-scale structure is close to becoming a reality. This is a necessary further step in the understanding of the cosmos, since the background evolution of the Universe seems to be in good agreement with the \lcdm\ paradigm. Therefore, the r\'egime of cosmological perturbations is the only arena where to detect deviations from it.

For those reasons, we have focussed on two viable $f(T)$ models. The first model, dubbed `exp', has been proven to be in good agreement with data concerning the background cosmic evolution, and its model parameter values have been accordingly fitted \citep{Cardone:2012xq}. The second model, baptised `pl', is a power-law generalisation of the teleparallel Lagrangian, and reduces to \lcdm\ when its power $n_T$ vanishes; we have therefore taken $n_T=0$ as fiducial. Regarding the experimental set-up, we have chosen the ESA \textit{Euclid} satellite as a reference survey. By doing so, we are able to perform both galaxy clustering measurements and cosmic shear tomography within the same experiment. This is utterly useful for our purpose, since the two probes are highly complementary and each one helps in lifting the other own degeneracies.

We have performed a Fisher matrix analysis to forecast the survey constraining potential and estimate the errors on parameter measurements. To better understand the most important aspects of the problem, we have firstly pursued galaxy clustering and cosmic shear alone. Respectively, forecast $1\sigma$ marginal errors on $f(T)$ model parameters are presented in Tables~\ref{tab:constraints-clustering3d}, \ref{tab:constraints-clustering2d} and \ref{tab:constraints-shear}, the second and third ones for three surveying configurations. Modified torsion gravity parameters are quite differently constrained by clustering and weak lensing. This is an interesting and novel result, and enables us to more deeply understand the properties of the $f(T)$ models under investigation. For example, constraints on extra-\lcdm\ parameters $n_T$ and $p_T$ of the $f(T)$-exp model are poorly constrained by three-dimensional clustering. This is because, over the redshift range $0.7-2.0$ probed by the \textit{Euclid} spectroscopic galaxy survey, the term $T_0/T$ in the exponential in Eq.~\eqref{eq:ftexp} quickly decreases. Hence, the $f(T)$ term in the Lagrangian becomes subdominant. On the contrary, the photometric imaging survey covers a wider redshift range, and this yields $5$ to $\sim10$ times tighter constraints on $n_T$, and $4$ to $>7$ times on $p_T$. This behaviour is reversed when analysing the $f(T)$-pl model: weak lensing error bounds are almost a order of magnitude broader than those from galaxy clustering. This somehow unexpected result may be explained by the strong degeneracies amongst the $n_T$ slope and, particularly, background-related \lcdm\ parameters.

The primary result of this work however comes from the combination of clustering and shear, presented in Sect.~\ref{ssec:res-tot}. The high complementarity of the two cosmological observables yields an impressive enhancement in the survey constraining power. This can be clearly seen in Fig.~\ref{fig:ellipses-tot}, where we show the forecast $1\sigma$ two-parameter marginal contours on $f(T)$-exp and pl model parameters, in the $(\vartheta_\alpha,\,\vartheta_\beta)$-planes. Light (darker) colour ellipses refer to galaxy clustering (cosmic shear), whereas the combination of the two \textit{Euclid} observables is depicted by the smallest, darkest ellipses. As a final result, we can quote the final $68.3\%$ marginal errors: $\sigma(n_T)=0.063$ and $\sigma(p_T)=0.14$, for the exp model, and $\sigma(n_T)=0.0097$, for the pl model.

Eventually, in Sect.~\ref{sec:lnB} we have made use of the reference \textit{Euclid} experiment as a tool for model selection. The calculation of the Bayes factor---the ratio of the posterior evidence probabilities---of two competing models allows the (predicted) data to decide whether one model is favoured over the other. In a sense, it provides us with a useful Bayesian Occam's razor that can assess which theoretical framework is preferred given the data---without computing the chi squared, for it will always reduce if we add more parameters. Within this framework, we compared the concordance \lcdm\ model and the $f(T)$-pl model; the former is in fact formally a subclass of the latter with $n_T=0$. Specifically, we have found that if \textit{Euclid} measure a non-zero value for $n_T$ of a few percent, there will be a strong evidence in favour of modified torsion gravity over \lcdm\ (see Fig.~\ref{fig:lnB}).

\acknowledgments We thank Pedro G. Ferreira for precious comments and Bianca Garilli for the spectroscopic redshift distribution of \textit{Euclid} galaxies. SC is funded by FCT-Portugal under Post-Doctoral Grant No. SFRH/BPD/80274/2011. VFC is funded by Italian Space Agency (ASI) through contract Euclid-IC (I/0.31/10/0). VFC acknowledges financial contribution from the agreement ASI/INAF/I/023/12/0. NR wishes to thank Agenzia Spaziale Italiana (ASI) for partial support. SC gratefully acknowledges Composita Creative Collective for hospitality during the development of this project.

\bibliographystyle{apsrev4-1}
\bibliography{/home/stefano/Documents/LaTeX/Bibliography}

%Merlin.mbs v4.21 2009-07-09.
\begin{thebibliography}{10}%
\makeatletter
\providecommand \@ifxundefined [1]{%
 \ifx #1\undefined \expandafter \@firstoftwo
 \else \expandafter \@secondoftwo
\fi
}%
\providecommand \@ifnum [1]{%
 \ifnum #1\expandafter \@firstoftwo
 \else \expandafter \@secondoftwo
\fi
}%
\providecommand \enquote [1]{``#1''}%
\providecommand \bibnamefont  [1]{#1}%
\providecommand \bibfnamefont [1]{#1}%
\providecommand \citenamefont [1]{#1}%
\providecommand\href[0]{\@sanitize\@href}%
\providecommand\@href[1]{\endgroup\@@startlink{#1}\endgroup\@@href}%
\providecommand\@@href[1]{#1\@@endlink}%
\providecommand \@sanitize [0]{\begingroup\catcode`\&12\catcode`\#12\relax}%
\@ifxundefined \pdfoutput {\@firstoftwo}{%
 \@ifnum{\z@=\pdfoutput}{\@firstoftwo}{\@secondoftwo}%
}{%
 \providecommand\@@startlink[1]{\leavevmode\special{html:<a href="#1">}}%
 \providecommand\@@endlink[0]{\special{html:</a>}}%
}{%
 \providecommand\@@startlink[1]{%
  \leavevmode
  \pdfstartlink
   attr{/Border[0 0 1 ]/H/I/C[0 1 1]}%
   user{/Subtype/Link/A<</Type/Action/S/URI/URI(#1)>>}%
  \relax
 }%
 \providecommand\@@endlink[0]{\pdfendlink}%
}%
\providecommand \url  [0]{\begingroup\@sanitize \@url }%
\providecommand \@url [1]{\endgroup\@href {#1}{\urlprefix}}%
\providecommand \urlprefix [0]{URL }%
\providecommand \Eprint[0]{\href }%
\@ifxundefined \urlstyle {%
  \providecommand \doi [1]{doi:\discretionary{}{}{}#1}%
}{%
  \providecommand \doi [0]{doi:\discretionary{}{}{}\begingroup
  \urlstyle{rm}\Url }%
}%
\providecommand \doibase [0]{http://dx.doi.org/}%
\providecommand \Doi[1]{\href{\doibase#1}}%
\providecommand \bibAnnote [3]{%
  \BibitemShut{#1}%
  \begin{quotation}\noindent
    \textsc{Key:}\ #2\\\textsc{Annotation:}\ #3%
  \end{quotation}%
}%
\providecommand \bibAnnoteFile [2]{%
  \IfFileExists{#2}{\bibAnnote {#1} {#2} {\input{#2}}}{}%
}%
\providecommand \typeout [0]{\immediate \write \m@ne }%
\providecommand \selectlanguage [0]{\@gobble}%
\providecommand \bibinfo [0]{\@secondoftwo}%
\providecommand \bibfield [0]{\@secondoftwo}%
\providecommand \translation [1]{[#1]}%
\providecommand \BibitemOpen[0]{}%
\providecommand \bibitemStop [0]{}%
\providecommand \bibitemNoStop [0]{.\EOS\space}%
\providecommand \EOS [0]{\spacefactor3000\relax}%
\providecommand \BibitemShut [1]{\csname bibitem#1\endcsname}%
%</preamble>
\bibitem{Riess:1998cb}%
  \BibitemOpen
  \bibfield{author}{%
  \bibinfo {author} {\bibfnamefont{A.~G.}\ \bibnamefont{Riess}} \emph{et~al.}
  (\bibinfo {collaboration} {Supernova Search Team}),\ }%
  \bibfield{journal}{%
  \Doi{10.1086/300499}{\bibinfo {journal} {Astron. J.}}\ }%
  \textbf{\bibinfo {volume} {116}},\ \bibinfo {pages} {1009} (\bibinfo {year}
  {1998}),\
  \Eprint{http://arxiv.org/abs/astro-ph/9805201}{arXiv:astro-ph/9805201}%
  \bibAnnoteFile{NoStop}{Riess:1998cb}%
%%CITATION = ASTRO-PH/9805201;%%
\bibitem{Perlmutter:1998np}%
  \BibitemOpen
  \bibfield{author}{%
  \bibinfo {author} {\bibfnamefont{S.}~\bibnamefont{Perlmutter}} \emph{et~al.}
  (\bibinfo {collaboration} {Supernova Cosmology Project}),\ }%
  \bibfield{journal}{%
  \Doi{10.1086/307221}{\bibinfo {journal} {Astrophys. J.}}\ }%
  \textbf{\bibinfo {volume} {517}},\ \bibinfo {pages} {565} (\bibinfo {year}
  {1999}),\
  \Eprint{http://arxiv.org/abs/astro-ph/9812133}{arXiv:astro-ph/9812133}%
  \bibAnnoteFile{NoStop}{Perlmutter:1998np}%
%%CITATION = ASTRO-PH/9812133;%%
\bibitem{Hinshaw:2012aka}%
  \BibitemOpen
  \bibfield{author}{%
  \bibinfo {author} {\bibfnamefont{G.}~\bibnamefont{Hinshaw}} \emph{et~al.}
  (\bibinfo {collaboration} {WMAP Collaboration})}%
   (\bibinfo {year} {2012}),\
  \Eprint{http://arxiv.org/abs/1212.5226}{arXiv:1212.5226 [astro-ph.CO]}%
  \bibAnnoteFile{NoStop}{Hinshaw:2012aka}%
%%CITATION = ARXIV:1212.5226;%%
\bibitem{Eisenstein:2005su}%
  \BibitemOpen
  \bibfield{author}{%
  \bibinfo {author} {\bibfnamefont{D.~J.}\ \bibnamefont{Eisenstein}}
  \emph{et~al.} (\bibinfo {collaboration} {SDSS}),\ }%
  \bibfield{journal}{%
  \Doi{10.1086/466512}{\bibinfo {journal} {Astrophys. J.}}\ }%
  \textbf{\bibinfo {volume} {633}},\ \bibinfo {pages} {560} (\bibinfo {year}
  {2005}),\
  \Eprint{http://arxiv.org/abs/astro-ph/0501171}{arXiv:astro-ph/0501171}%
  \bibAnnoteFile{NoStop}{Eisenstein:2005su}%
%%CITATION = ASTRO-PH/0501171;%%
\bibitem{Wang:2008vja}%
  \BibitemOpen
  \bibfield{author}{%
  \bibinfo {author} {\bibfnamefont{Y.}~\bibnamefont{Wang}},\ }%
  \bibfield{journal}{%
  \Doi{10.1103/PhysRevD.78.123532}{\bibinfo {journal} {Phys. Rev.}}\ }%
  \textbf{\bibinfo {volume} {D78}},\ \bibinfo {pages} {123532} (\bibinfo {year}
  {2008}),\ \Eprint{http://arxiv.org/abs/0809.0657}{arXiv:0809.0657
  [astro-ph]}%
  \bibAnnoteFile{NoStop}{Wang:2008vja}%
%%CITATION = ARXIV:0809.0657;%%
\bibitem{Weinberg:1988cp}%
  \BibitemOpen
  \bibfield{author}{%
  \bibinfo {author} {\bibfnamefont{S.}~\bibnamefont{Weinberg}},\ }%
  \bibfield{journal}{%
  \Doi{10.1103/RevModPhys.61.1}{\bibinfo {journal} {Rev. Mod. Phys.}}\ }%
  \textbf{\bibinfo {volume} {61}},\ \bibinfo {pages} {1} (\bibinfo {year}
  {1989}),\ \bibinfo {note} {morris Loeb Lectures in Physics, Harvard
  University, May 2, 3, 5, and 10, 1988}%
  \bibAnnoteFile{NoStop}{Weinberg:1988cp}%
\bibitem{Einstein:28}%
  \BibitemOpen
  \bibfield{author}{%
  \bibinfo {author} {\bibfnamefont{A.}~\bibnamefont{Einstein}},\ }%
  \bibfield{journal}{%
  \bibinfo {journal} {Sitzungsber. Preuss. Akad. Wiss. Phys. Math. KI.},\
  \bibinfo {pages} {217}}%
   (\bibinfo {year} {1928})%
  \bibAnnoteFile{NoStop}{Einstein:28}%
\bibitem{Einstein:30a}%
  \BibitemOpen
  \bibfield{author}{%
  \bibinfo {author} {\bibfnamefont{A.}~\bibnamefont{Einstein}},\ }%
  \bibfield{journal}{%
  \bibinfo {journal} {Sitzungsber. Preuss. Akad. Wiss. Phys. Math. KI.},\
  \bibinfo {pages} {401}}%
   (\bibinfo {year} {1930})%
  \bibAnnoteFile{NoStop}{Einstein:30a}%
\bibitem{Einstein:30b}%
  \BibitemOpen
  \bibfield{author}{%
  \bibinfo {author} {\bibfnamefont{A.}~\bibnamefont{Einstein}},\ }%
  \bibfield{journal}{%
  \bibinfo {journal} {Math. Annal.}\ }%
  \textbf{\bibinfo {volume} {102}},\ \bibinfo {pages} {65} (\bibinfo {year}
  {1930})%
  \bibAnnoteFile{NoStop}{Einstein:30b}%
\bibitem{Arcos:2005ec}%
  \BibitemOpen
  \bibfield{author}{%
  \bibinfo {author} {\bibfnamefont{H.~I.}\ \bibnamefont{Arcos}}\ and\ \bibinfo
  {author} {\bibfnamefont{J.~G.}\ \bibnamefont{Pereira}},\ }%
  \bibfield{journal}{%
  \Doi{10.1142/S0218271804006462}{\bibinfo {journal} {Int. J. Mod. Phys.}}\ }%
  \textbf{\bibinfo {volume} {D13}},\ \bibinfo {pages} {2193} (\bibinfo {year}
  {2004}),\ \Eprint{http://arxiv.org/abs/gr-qc/0501017}{arXiv:gr-qc/0501017}%
  \bibAnnoteFile{NoStop}{Arcos:2005ec}%
%%CITATION = GR-QC/0501017;%%
\bibitem{Capozziello:2003tk}%
  \BibitemOpen
  \bibfield{author}{%
  \bibinfo {author} {\bibfnamefont{S.}~\bibnamefont{Capozziello}}, \bibinfo
  {author} {\bibfnamefont{S.}~\bibnamefont{Carloni}},\ and\ \bibinfo {author}
  {\bibfnamefont{A.}~\bibnamefont{Troisi}},\ }%
  \bibfield{journal}{%
  \bibinfo {journal} {Recent Res. Dev. Astron. Astrophys.}\ }%
  \textbf{\bibinfo {volume} {1}},\ \bibinfo {pages} {625} (\bibinfo {year}
  {2003}),\
  \Eprint{http://arxiv.org/abs/astro-ph/0303041}{arXiv:astro-ph/0303041
  [astro-ph]}%
  \bibAnnoteFile{NoStop}{Capozziello:2003tk}%
%%CITATION = ASTRO-PH/0303041;%%
\bibitem{Carroll:2003wy}%
  \BibitemOpen
  \bibfield{author}{%
  \bibinfo {author} {\bibfnamefont{S.~M.}\ \bibnamefont{Carroll}}, \bibinfo
  {author} {\bibfnamefont{V.}~\bibnamefont{Duvvuri}}, \bibinfo {author}
  {\bibfnamefont{M.}~\bibnamefont{Trodden}},\ and\ \bibinfo {author}
  {\bibfnamefont{M.~S.}\ \bibnamefont{Turner}},\ }%
  \bibfield{journal}{%
  \Doi{10.1103/PhysRevD.70.043528}{\bibinfo {journal} {Phys. Rev.}}\ }%
  \textbf{\bibinfo {volume} {D70}},\ \bibinfo {pages} {043528} (\bibinfo {year}
  {2004}),\
  \Eprint{http://arxiv.org/abs/astro-ph/0306438}{arXiv:astro-ph/0306438}%
  \bibAnnoteFile{NoStop}{Carroll:2003wy}%
%%CITATION = ASTRO-PH/0306438;%%
\bibitem{Ferraro:2008ey}%
  \BibitemOpen
  \bibfield{author}{%
  \bibinfo {author} {\bibfnamefont{R.}~\bibnamefont{Ferraro}}\ and\ \bibinfo
  {author} {\bibfnamefont{F.}~\bibnamefont{Fiorini}},\ }%
  \bibfield{journal}{%
  \Doi{10.1103/PhysRevD.78.124019}{\bibinfo {journal} {Phys. Rev.}}\ }%
  \textbf{\bibinfo {volume} {D78}},\ \bibinfo {pages} {124019} (\bibinfo {year}
  {2008}),\ \Eprint{http://arxiv.org/abs/0812.1981}{arXiv:0812.1981 [gr-qc]}%
  \bibAnnoteFile{NoStop}{Ferraro:2008ey}%
%%CITATION = ARXIV:0812.1981;%%
\bibitem{Fiorini:2009ux}%
  \BibitemOpen
  \bibfield{author}{%
  \bibinfo {author} {\bibfnamefont{F.}~\bibnamefont{Fiorini}}\ and\ \bibinfo
  {author} {\bibfnamefont{R.}~\bibnamefont{Ferraro}},\ }%
  \bibfield{journal}{%
  \Doi{10.1142/S0217751X09045236}{\bibinfo {journal} {Int. J. Mod. Phys.}}\ }%
  \textbf{\bibinfo {volume} {A24}},\ \bibinfo {pages} {1686} (\bibinfo {year}
  {2009}),\ \Eprint{http://arxiv.org/abs/0904.1767}{arXiv:0904.1767 [gr-qc]}%
  \bibAnnoteFile{NoStop}{Fiorini:2009ux}%
%%CITATION = ARXIV:0904.1767;%%
\bibitem{Li:2010cg}%
  \BibitemOpen
  \bibfield{author}{%
  \bibinfo {author} {\bibfnamefont{B.}~\bibnamefont{Li}}, \bibinfo {author}
  {\bibfnamefont{T.~P.}\ \bibnamefont{Sotiriou}},\ and\ \bibinfo {author}
  {\bibfnamefont{J.~D.}\ \bibnamefont{Barrow}},\ }%
  \bibfield{journal}{%
  \Doi{10.1103/PhysRevD.83.064035}{\bibinfo {journal} {Phys. Rev.}}\ }%
  \textbf{\bibinfo {volume} {D83}},\ \bibinfo {pages} {064035} (\bibinfo {year}
  {2011}),\ \Eprint{http://arxiv.org/abs/1010.1041}{arXiv:1010.1041 [gr-qc]}%
  \bibAnnoteFile{NoStop}{Li:2010cg}%
%%CITATION = ARXIV:1010.1041;%%
\bibitem{2010deto.book.....A}%
  \BibitemOpen
  \bibfield{author}{%
  \bibinfo {author} {\bibfnamefont{L.}~\bibnamefont{{Amendola}}}\ and\ \bibinfo
  {author} {\bibfnamefont{S.}~\bibnamefont{{Tsujikawa}}},\ }%
  \emph{\bibinfo {title} {{Dark Energy: Theory and Observations}}}\ (\bibinfo
  {year} {2010})\ \bibinfo {note} {~Cambridge University Press}%
  \bibAnnoteFile{NoStop}{2010deto.book.....A}%
\bibitem{DeFelice:2010aj}%
  \BibitemOpen
  \bibfield{author}{%
  \bibinfo {author} {\bibfnamefont{A.}~\bibnamefont{De~Felice}}\ and\ \bibinfo
  {author} {\bibfnamefont{S.}~\bibnamefont{Tsujikawa}},\ }%
  \bibfield{journal}{%
  \bibinfo {journal} {Living Rev. Rel.}\ }%
  \textbf{\bibinfo {volume} {13}},\ \bibinfo {pages} {3} (\bibinfo {year}
  {2010}),\ \Eprint{http://arxiv.org/abs/1002.4928}{arXiv:1002.4928 [gr-qc]}%
  \bibAnnoteFile{NoStop}{DeFelice:2010aj}%
%%CITATION = ARXIV:1002.4928;%%
\bibitem{Nojiri:2005vv}%
  \BibitemOpen
  \bibfield{author}{%
  \bibinfo {author} {\bibfnamefont{S.}~\bibnamefont{Nojiri}}, \bibinfo {author}
  {\bibfnamefont{S.~D.}\ \bibnamefont{Odintsov}},\ and\ \bibinfo {author}
  {\bibfnamefont{M.}~\bibnamefont{Sasaki}},\ }%
  \bibfield{journal}{%
  \Doi{10.1103/PhysRevD.71.123509}{\bibinfo {journal} {Phys. Rev.}}\ }%
  \textbf{\bibinfo {volume} {D71}},\ \bibinfo {pages} {123509} (\bibinfo {year}
  {2005}),\ \Eprint{http://arxiv.org/abs/hep-th/0504052}{arXiv:hep-th/0504052
  [hep-th]}%
  \bibAnnoteFile{NoStop}{Nojiri:2005vv}%
%%CITATION = HEP-TH/0504052;%%
\bibitem{Koivisto:2006xf}%
  \BibitemOpen
  \bibfield{author}{%
  \bibinfo {author} {\bibfnamefont{T.}~\bibnamefont{Koivisto}}\ and\ \bibinfo
  {author} {\bibfnamefont{D.~F.}\ \bibnamefont{Mota}},\ }%
  \bibfield{journal}{%
  \Doi{10.1016/j.physletb.2006.11.048}{\bibinfo {journal} {Phys. Lett.}}\ }%
  \textbf{\bibinfo {volume} {B644}},\ \bibinfo {pages} {104} (\bibinfo {year}
  {2007}),\
  \Eprint{http://arxiv.org/abs/astro-ph/0606078}{arXiv:astro-ph/0606078}%
  \bibAnnoteFile{NoStop}{Koivisto:2006xf}%
%%CITATION = ASTRO-PH/0606078;%%
\bibitem{Brans:1961sx}%
  \BibitemOpen
  \bibfield{author}{%
  \bibinfo {author} {\bibfnamefont{C.}~\bibnamefont{Brans}}\ and\ \bibinfo
  {author} {\bibfnamefont{R.}~\bibnamefont{Dicke}},\ }%
  \bibfield{journal}{%
  \Doi{10.1103/PhysRev.124.925}{\bibinfo {journal} {Phys. Rev.}}\ }%
  \textbf{\bibinfo {volume} {124}},\ \bibinfo {pages} {925} (\bibinfo {year}
  {1961})%
  \bibAnnoteFile{NoStop}{Brans:1961sx}%
%%CITATION = PHRVA,124,925;%%
\bibitem{EspositoFarese:2000ij}%
  \BibitemOpen
  \bibfield{author}{%
  \bibinfo {author} {\bibfnamefont{G.}~\bibnamefont{Esposito-Farese}}\ and\
  \bibinfo {author} {\bibfnamefont{D.}~\bibnamefont{Polarski}},\ }%
  \bibfield{journal}{%
  \Doi{10.1103/PhysRevD.63.063504}{\bibinfo {journal} {Phys. Rev.}}\ }%
  \textbf{\bibinfo {volume} {D63}},\ \bibinfo {pages} {063504} (\bibinfo {year}
  {2001}),\ \Eprint{http://arxiv.org/abs/gr-qc/0009034}{arXiv:gr-qc/0009034
  [gr-qc]}%
  \bibAnnoteFile{NoStop}{EspositoFarese:2000ij}%
%%CITATION = GR-QC/0009034;%%
\bibitem{ArmendarizPicon:1999rj}%
  \BibitemOpen
  \bibfield{author}{%
  \bibinfo {author} {\bibfnamefont{C.}~\bibnamefont{Armendariz-Picon}},
  \bibinfo {author} {\bibfnamefont{T.}~\bibnamefont{Damour}},\ and\ \bibinfo
  {author} {\bibfnamefont{V.~F.}\ \bibnamefont{Mukhanov}},\ }%
  \bibfield{journal}{%
  \Doi{10.1016/S0370-2693(99)00603-6}{\bibinfo {journal} {Phys. Lett.}}\ }%
  \textbf{\bibinfo {volume} {B458}},\ \bibinfo {pages} {209} (\bibinfo {year}
  {1999}),\ \Eprint{http://arxiv.org/abs/hep-th/9904075}{arXiv:hep-th/9904075}%
  \bibAnnoteFile{NoStop}{ArmendarizPicon:1999rj}%
%%CITATION = HEP-TH/9904075;%%
\bibitem{ArmendarizPicon:2000ah}%
  \BibitemOpen
  \bibfield{author}{%
  \bibinfo {author} {\bibfnamefont{C.}~\bibnamefont{Armendariz-Picon}},
  \bibinfo {author} {\bibfnamefont{V.~F.}\ \bibnamefont{Mukhanov}},\ and\
  \bibinfo {author} {\bibfnamefont{P.~J.}\ \bibnamefont{Steinhardt}},\ }%
  \bibfield{journal}{%
  \Doi{10.1103/PhysRevD.63.103510}{\bibinfo {journal} {Phys. Rev.}}\ }%
  \textbf{\bibinfo {volume} {D63}},\ \bibinfo {pages} {103510} (\bibinfo {year}
  {2001}),\
  \Eprint{http://arxiv.org/abs/astro-ph/0006373}{arXiv:astro-ph/0006373}%
  \bibAnnoteFile{NoStop}{ArmendarizPicon:2000ah}%
%%CITATION = ASTRO-PH/0006373;%%
\bibitem{Nicolis:2008in}%
  \BibitemOpen
  \bibfield{author}{%
  \bibinfo {author} {\bibfnamefont{A.}~\bibnamefont{Nicolis}}, \bibinfo
  {author} {\bibfnamefont{R.}~\bibnamefont{Rattazzi}},\ and\ \bibinfo {author}
  {\bibfnamefont{E.}~\bibnamefont{Trincherini}},\ }%
  \bibfield{journal}{%
  \Doi{10.1103/PhysRevD.79.064036}{\bibinfo {journal} {Phys. Rev.}}\ }%
  \textbf{\bibinfo {volume} {D79}},\ \bibinfo {pages} {064036} (\bibinfo {year}
  {2009}),\ \Eprint{http://arxiv.org/abs/0811.2197}{arXiv:0811.2197 [hep-th]}%
  \bibAnnoteFile{NoStop}{Nicolis:2008in}%
%%CITATION = ARXIV:0811.2197;%%
\bibitem{Burrage:2011bt}%
  \BibitemOpen
  \bibfield{author}{%
  \bibinfo {author} {\bibfnamefont{C.}~\bibnamefont{Burrage}}, \bibinfo
  {author} {\bibfnamefont{C.}~\bibnamefont{de~Rham}},\ and\ \bibinfo {author}
  {\bibfnamefont{L.}~\bibnamefont{Heisenberg}},\ }%
  \bibfield{journal}{%
  \Doi{10.1088/1475-7516/2011/05/025}{\bibinfo {journal} {JCAP}}\ }%
  \textbf{\bibinfo {volume} {1105}},\ \bibinfo {pages} {025} (\bibinfo {year}
  {2011}),\ \Eprint{http://arxiv.org/abs/1104.0155}{arXiv:1104.0155 [hep-th]}%
  \bibAnnoteFile{NoStop}{Burrage:2011bt}%
%%CITATION = ARXIV:1104.0155;%%
\bibitem{Gao:2011mz}%
  \BibitemOpen
  \bibfield{author}{%
  \bibinfo {author} {\bibfnamefont{X.}~\bibnamefont{Gao}},\ }%
  \bibfield{journal}{%
  \Doi{10.1088/1475-7516/2011/10/021}{\bibinfo {journal} {JCAP}}\ }%
  \textbf{\bibinfo {volume} {1110}},\ \bibinfo {pages} {021} (\bibinfo {year}
  {2011}),\ \Eprint{http://arxiv.org/abs/1106.0292}{arXiv:1106.0292
  [astro-ph.CO]}%
  \bibAnnoteFile{NoStop}{Gao:2011mz}%
%%CITATION = ARXIV:1106.0292;%%
\bibitem{Cardone:2012xq}%
  \BibitemOpen
  \bibfield{author}{%
  \bibinfo {author} {\bibfnamefont{V.~F.}\ \bibnamefont{Cardone}}, \bibinfo
  {author} {\bibfnamefont{N.}~\bibnamefont{Radicella}},\ and\ \bibinfo {author}
  {\bibfnamefont{S.}~\bibnamefont{Camera}},\ }%
  \bibfield{journal}{%
  \Doi{10.1103/PhysRevD.85.124007}{\bibinfo {journal} {Phys. Rev.}}\ }%
  \textbf{\bibinfo {volume} {D85}},\ \bibinfo {pages} {124007} (\bibinfo {year}
  {2012}),\ \Eprint{http://arxiv.org/abs/1204.5294}{arXiv:1204.5294
  [astro-ph.CO]}%
  \bibAnnoteFile{NoStop}{Cardone:2012xq}%
%%CITATION = ARXIV:1204.5294;%%
\bibitem{Abbott:2005bi}%
  \BibitemOpen
  \bibfield{author}{%
  \bibinfo {author} {\bibfnamefont{T.}~\bibnamefont{Abbott}} \emph{et~al.}
  (\bibinfo {collaboration} {Dark Energy Survey})}%
   (\bibinfo {year} {2005}),\
  \Eprint{http://arxiv.org/abs/astro-ph/0510346}{arXiv:astro-ph/0510346}%
  \bibAnnoteFile{NoStop}{Abbott:2005bi}%
%%CITATION = ASTRO-PH/0510346;%%
\bibitem{EditorialTeam:2011mu}%
  \BibitemOpen
  \bibfield{author}{%
  \bibinfo {author} {\bibfnamefont{R.}~\bibnamefont{Laureijs}}, \bibinfo
  {author} {\bibfnamefont{J.}~\bibnamefont{Amiaux}}, \bibinfo {author}
  {\bibfnamefont{S.}~\bibnamefont{Arduini}}, \bibinfo {author}
  {\bibfnamefont{J.-L.}\ \bibnamefont{Augueres}}, \emph{et~al.} (\bibinfo
  {collaboration} {Euclid}),\ }%
  \bibfield{journal}{%
  \bibinfo {journal} {ESA-SRE}\ }%
  \textbf{\bibinfo {volume} {12}} (\bibinfo {year} {2011}),\
  \Eprint{http://arxiv.org/abs/1110.3193}{arXiv:1110.3193 [astro-ph.CO]}%
  \bibAnnoteFile{NoStop}{EditorialTeam:2011mu}%
\bibitem{Amendola:2012ys}%
  \BibitemOpen
  \bibfield{author}{%
  \bibinfo {author} {\bibfnamefont{L.}~\bibnamefont{Amendola}} \emph{et~al.}
  (\bibinfo {collaboration} {Euclid Theory Working Group}),\ }%
  \bibfield{journal}{%
  \bibinfo {journal} {Living Rev.Rel.}\ }%
  \textbf{\bibinfo {volume} {16}},\ \bibinfo {pages} {6} (\bibinfo {year}
  {2013}),\ \Eprint{http://arxiv.org/abs/1206.1225}{arXiv:1206.1225
  [astro-ph.CO]}%
  \bibAnnoteFile{NoStop}{Amendola:2012ys}%
%%CITATION = ARXIV:1206.1225;%%
\bibitem{Ivezic:2008fe}%
  \BibitemOpen
  \bibfield{author}{%
  \bibinfo {author} {\bibfnamefont{Z.}~\bibnamefont{Ivezic}}, \bibinfo {author}
  {\bibfnamefont{J.}~\bibnamefont{Tyson}}, \bibinfo {author}
  {\bibfnamefont{R.}~\bibnamefont{Allsman}}, \bibinfo {author}
  {\bibfnamefont{J.}~\bibnamefont{Andrew}}, \bibinfo {author}
  {\bibfnamefont{R.}~\bibnamefont{Angel}}, \emph{et~al.}}%
   (\bibinfo {year} {2008}),\
  \Eprint{http://arxiv.org/abs/0805.2366}{arXiv:0805.2366 [astro-ph]}%
  \bibAnnoteFile{NoStop}{Ivezic:2008fe}%
%%CITATION = ARXIV:0805.2366;%%
\bibitem{Abate:2012za}%
  \BibitemOpen
  \bibfield{author}{%
  \bibinfo {author} {\bibfnamefont{A.}~\bibnamefont{Abate}} \emph{et~al.}
  (\bibinfo {collaboration} {LSST Dark Energy Science Collaboration})}%
   (\bibinfo {year} {2012}),\
  \Eprint{http://arxiv.org/abs/1211.0310}{arXiv:1211.0310 [astro-ph.CO]}%
  \bibAnnoteFile{NoStop}{Abate:2012za}%
%%CITATION = ARXIV:1211.0310;%%
\bibitem{2013IAUS..291..337T}%
  \BibitemOpen
  \bibfield{author}{%
  \bibinfo {author} {\bibfnamefont{A.~R.}\ \bibnamefont{{Taylor}}},\ }%
  \bibinfo {series} {IAU Symposium}\ \textbf{\bibinfo {volume} {291}},\
  \bibinfo {pages} {337} (\bibinfo {year} {2013})%
  \bibAnnoteFile{NoStop}{2013IAUS..291..337T}%
\bibitem{2008ExA....22..151J}%
  \BibitemOpen
  \bibfield{author}{%
  \bibinfo {author} {\bibfnamefont{S.}~\bibnamefont{{Johnston}}}, \bibinfo
  {author} {\bibfnamefont{R.}~\bibnamefont{{Taylor}}}, \bibinfo {author}
  {\bibfnamefont{M.}~\bibnamefont{{Bailes}}}, \bibinfo {author}
  {\bibfnamefont{N.}~\bibnamefont{{Bartel}}}, \bibinfo {author}
  {\bibfnamefont{C.}~\bibnamefont{{Baugh}}}, \emph{et~al.},\ }%
  \bibfield{journal}{%
  \Doi{10.1007/s10686-008-9124-7}{\bibinfo {journal} {Experimental Astronomy}}\
  }%
  \textbf{\bibinfo {volume} {22}},\ \bibinfo {pages} {151} (\bibinfo {month}
  {Dec.}\ \bibinfo {year} {2008}),\
  \Eprint{http://arxiv.org/abs/0810.5187}{arXiv:0810.5187}%
  \bibAnnoteFile{NoStop}{2008ExA....22..151J}%
\bibitem{Oosterloo:2010wz}%
  \BibitemOpen
  \bibfield{author}{%
  \bibinfo {author} {\bibfnamefont{T.}~\bibnamefont{Oosterloo}}, \bibinfo
  {author} {\bibfnamefont{M.}~\bibnamefont{Verheijen}},\ and\ \bibinfo {author}
  {\bibfnamefont{W.}~\bibnamefont{van Cappellen}},\ }%
  \bibfield{journal}{%
  \bibinfo {journal} {PoS}\ }%
  \textbf{\bibinfo {volume} {ISKAF2010}},\ \bibinfo {pages} {043} (\bibinfo
  {year} {2010}),\ \Eprint{http://arxiv.org/abs/1007.5141}{arXiv:1007.5141
  [astro-ph.IM]}%
  \bibAnnoteFile{NoStop}{Oosterloo:2010wz}%
\bibitem{Norris:2011ai}%
  \BibitemOpen
  \bibfield{author}{%
  \bibinfo {author} {\bibfnamefont{R.~P.}\ \bibnamefont{{Norris}}}, \bibinfo
  {author} {\bibfnamefont{A.~M.}\ \bibnamefont{{Hopkins}}}, \bibinfo {author}
  {\bibfnamefont{J.}~\bibnamefont{{Afonso}}}, \bibinfo {author}
  {\bibfnamefont{S.}~\bibnamefont{{Brown}}}, \bibinfo {author}
  {\bibfnamefont{J.~J.}\ \bibnamefont{{Condon}}}, \emph{et~al.},\ }%
  \bibfield{journal}{%
  \Doi{10.1071/AS11021}{\bibinfo {journal} {PASA}}\ }%
  \textbf{\bibinfo {volume} {28}},\ \bibinfo {pages} {215} (\bibinfo {month}
  {Aug.}\ \bibinfo {year} {2011}),\
  \Eprint{http://arxiv.org/abs/1106.3219}{arXiv:1106.3219 [astro-ph.CO]}%
  \bibAnnoteFile{NoStop}{Norris:2011ai}%
\bibitem{Rottgering:2011jq}%
  \BibitemOpen
  \bibfield{author}{%
  \bibinfo {author} {\bibfnamefont{H.}~\bibnamefont{Rottgering}}, \bibinfo
  {author} {\bibfnamefont{J.}~\bibnamefont{Afonso}}, \bibinfo {author}
  {\bibfnamefont{P.}~\bibnamefont{Barthel}}, \bibinfo {author}
  {\bibfnamefont{F.}~\bibnamefont{Batejat}}, \bibinfo {author}
  {\bibfnamefont{P.}~\bibnamefont{Best}}, \emph{et~al.},\ }%
  \bibfield{journal}{%
  \Doi{10.1007/s12036-011-9129-x}{\bibinfo {journal} {J. Astropy. and
  Astron.}}\ }%
  \textbf{\bibinfo {volume} {32}},\ \bibinfo {pages} {557} (\bibinfo {year}
  {2011}),\ \Eprint{http://arxiv.org/abs/1107.1606}{arXiv:1107.1606
  [astro-ph.CO]}%
  \bibAnnoteFile{NoStop}{Rottgering:2011jq}%
\bibitem{Trotta:2005ar}%
  \BibitemOpen
  \bibfield{author}{%
  \bibinfo {author} {\bibfnamefont{R.}~\bibnamefont{Trotta}},\ }%
  \bibfield{journal}{%
  \Doi{10.1111/j.1365-2966.2007.11738.x}{\bibinfo {journal} {Mon. Not. Roy.
  Astron. Soc.}}\ }%
  \textbf{\bibinfo {volume} {378}},\ \bibinfo {pages} {72} (\bibinfo {year}
  {2007}),\
  \Eprint{http://arxiv.org/abs/astro-ph/0504022}{arXiv:astro-ph/0504022}%
  \bibAnnoteFile{NoStop}{Trotta:2005ar}%
%%CITATION = ASTRO-PH/0504022;%%
\bibitem{Heavens:2007ka}%
  \BibitemOpen
  \bibfield{author}{%
  \bibinfo {author} {\bibfnamefont{A.~F.}\ \bibnamefont{Heavens}}, \bibinfo
  {author} {\bibfnamefont{T.~D.}\ \bibnamefont{Kitching}},\ and\ \bibinfo
  {author} {\bibfnamefont{L.}~\bibnamefont{Verde}},\ }%
  \bibfield{journal}{%
  \Doi{10.1111/j.1365-2966.2007.12134.x}{\bibinfo {journal} {Mon. Not. Roy.
  Astron. Soc.}}\ }%
  \textbf{\bibinfo {volume} {380}},\ \bibinfo {pages} {1029} (\bibinfo {year}
  {2007}),\
  \Eprint{http://arxiv.org/abs/astro-ph/0703191}{arXiv:astro-ph/0703191}%
  \bibAnnoteFile{NoStop}{Heavens:2007ka}%
%%CITATION = ASTRO-PH/0703191;%%
\bibitem{Camera:2011ms}%
  \BibitemOpen
  \bibfield{author}{%
  \bibinfo {author} {\bibfnamefont{S.}~\bibnamefont{Camera}}, \bibinfo {author}
  {\bibfnamefont{A.}~\bibnamefont{Diaferio}},\ and\ \bibinfo {author}
  {\bibfnamefont{V.~F.}\ \bibnamefont{Cardone}},\ }%
  \bibfield{journal}{%
  \Doi{10.1088/1475-7516/2011/07/016}{\bibinfo {journal} {JCAP}}\ }%
  \textbf{\bibinfo {volume} {1107}},\ \bibinfo {pages} {016} (\bibinfo {year}
  {2011}),\ \Eprint{http://arxiv.org/abs/1104.2740}{arXiv:1104.2740
  [astro-ph.CO]}%
  \bibAnnoteFile{NoStop}{Camera:2011ms}%
\bibitem{Camera:2010wm}%
  \BibitemOpen
  \bibfield{author}{%
  \bibinfo {author} {\bibfnamefont{S.}~\bibnamefont{Camera}}, \bibinfo {author}
  {\bibfnamefont{T.~D.}\ \bibnamefont{Kitching}}, \bibinfo {author}
  {\bibfnamefont{A.~F.}\ \bibnamefont{Heavens}}, \bibinfo {author}
  {\bibfnamefont{D.}~\bibnamefont{Bertacca}},\ and\ \bibinfo {author}
  {\bibfnamefont{A.}~\bibnamefont{Diaferio}},\ }%
  \bibfield{journal}{%
  \bibinfo {journal} {Mon. Not. Roy. Astron. Soc.}\ }%
  \textbf{\bibinfo {volume} {415}},\ \bibinfo {pages} {399} (\bibinfo {year}
  {2011}),\ \Eprint{http://arxiv.org/abs/1002.4740}{arXiv:1002.4740
  [astro-ph.CO]}%
  \bibAnnoteFile{NoStop}{Camera:2010wm}%
%%CITATION = 1002.4740;%%
\bibitem{Linder:2010py}%
  \BibitemOpen
  \bibfield{author}{%
  \bibinfo {author} {\bibfnamefont{E.~V.}\ \bibnamefont{Linder}},\ }%
  \bibfield{journal}{%
  \Doi{10.1103/PhysRevD.81.127301, 10.1103/PhysRevD.82.109902}{\bibinfo
  {journal} {Phys. Rev.}}\ }%
  \textbf{\bibinfo {volume} {D81}},\ \bibinfo {pages} {127301} (\bibinfo {year}
  {2010}),\ \Eprint{http://arxiv.org/abs/1005.3039}{arXiv:1005.3039
  [astro-ph.CO]}%
  \bibAnnoteFile{NoStop}{Linder:2010py}%
%%CITATION = ARXIV:1005.3039;%%
\bibitem{Wu:2010mn}%
  \BibitemOpen
  \bibfield{author}{%
  \bibinfo {author} {\bibfnamefont{P.}~\bibnamefont{Wu}}\ and\ \bibinfo
  {author} {\bibfnamefont{H.~W.}\ \bibnamefont{Yu}},\ }%
  \bibfield{journal}{%
  \Doi{10.1016/j.physletb.2010.08.073}{\bibinfo {journal} {Phys. Lett.}}\ }%
  \textbf{\bibinfo {volume} {B693}},\ \bibinfo {pages} {415} (\bibinfo {year}
  {2010}),\ \Eprint{http://arxiv.org/abs/1006.0674}{arXiv:1006.0674 [gr-qc]}%
  \bibAnnoteFile{NoStop}{Wu:2010mn}%
%%CITATION = ARXIV:1006.0674;%%
\bibitem{Bengochea:2010sg}%
  \BibitemOpen
  \bibfield{author}{%
  \bibinfo {author} {\bibfnamefont{G.~R.}\ \bibnamefont{Bengochea}},\ }%
  \bibfield{journal}{%
  \Doi{10.1016/j.physletb.2010.11.064}{\bibinfo {journal} {Phys. Lett.}}\ }%
  \textbf{\bibinfo {volume} {B695}},\ \bibinfo {pages} {405} (\bibinfo {year}
  {2011}),\ \Eprint{http://arxiv.org/abs/1008.3188}{arXiv:1008.3188
  [astro-ph.CO]}%
  \bibAnnoteFile{NoStop}{Bengochea:2010sg}%
%%CITATION = ARXIV:1008.3188;%%
\bibitem{Ferraro:2011us}%
  \BibitemOpen
  \bibfield{author}{%
  \bibinfo {author} {\bibfnamefont{R.}~\bibnamefont{Ferraro}}\ and\ \bibinfo
  {author} {\bibfnamefont{F.}~\bibnamefont{Fiorini}},\ }%
  \bibfield{journal}{%
  \Doi{10.1016/j.physletb.2011.06.049}{\bibinfo {journal} {Phys. Lett.}}\ }%
  \textbf{\bibinfo {volume} {B702}},\ \bibinfo {pages} {75} (\bibinfo {year}
  {2011}),\ \Eprint{http://arxiv.org/abs/1103.0824}{arXiv:1103.0824 [gr-qc]}%
  \bibAnnoteFile{NoStop}{Ferraro:2011us}%
%%CITATION = ARXIV:1103.0824;%%
\bibitem{Bengochea:2008gz}%
  \BibitemOpen
  \bibfield{author}{%
  \bibinfo {author} {\bibfnamefont{G.~R.}\ \bibnamefont{Bengochea}}\ and\
  \bibinfo {author} {\bibfnamefont{R.}~\bibnamefont{Ferraro}},\ }%
  \bibfield{journal}{%
  \Doi{10.1103/PhysRevD.79.124019}{\bibinfo {journal} {Phys. Rev.}}\ }%
  \textbf{\bibinfo {volume} {D79}},\ \bibinfo {pages} {124019} (\bibinfo {year}
  {2009}),\ \Eprint{http://arxiv.org/abs/0812.1205}{arXiv:0812.1205
  [astro-ph]}%
  \bibAnnoteFile{NoStop}{Bengochea:2008gz}%
%%CITATION = 0812.1205;%%
\bibitem{Zheng:2010am}%
  \BibitemOpen
  \bibfield{author}{%
  \bibinfo {author} {\bibfnamefont{R.}~\bibnamefont{Zheng}}\ and\ \bibinfo
  {author} {\bibfnamefont{Q.-G.}\ \bibnamefont{Huang}},\ }%
  \bibfield{journal}{%
  \Doi{10.1088/1475-7516/2011/03/002}{\bibinfo {journal} {JCAP}}\ }%
  \textbf{\bibinfo {volume} {1103}},\ \bibinfo {pages} {002} (\bibinfo {year}
  {2011}),\ \Eprint{http://arxiv.org/abs/1010.3512}{arXiv:1010.3512 [gr-qc]}%
  \bibAnnoteFile{NoStop}{Zheng:2010am}%
%%CITATION = ARXIV:1010.3512;%%
\bibitem{Fisher:1935}%
  \BibitemOpen
  \bibfield{author}{%
  \bibinfo {author} {\bibfnamefont{R.~A.}\ \bibnamefont{Fisher}},\ }%
  \bibfield{journal}{%
  \bibinfo {journal} {J. Roy. Stat. Soc.}\ }%
  \textbf{\bibinfo {volume} {98}},\ \bibinfo {pages} {39} (\bibinfo {year}
  {1935})%
  \bibAnnoteFile{NoStop}{Fisher:1935}%
\bibitem{Jungman:1995bz}%
  \BibitemOpen
  \bibfield{author}{%
  \bibinfo {author} {\bibfnamefont{G.}~\bibnamefont{Jungman}}, \bibinfo
  {author} {\bibfnamefont{M.}~\bibnamefont{Kamionkowski}}, \bibinfo {author}
  {\bibfnamefont{A.}~\bibnamefont{Kosowsky}},\ and\ \bibinfo {author}
  {\bibfnamefont{D.~N.}\ \bibnamefont{Spergel}},\ }%
  \bibfield{journal}{%
  \Doi{10.1103/PhysRevD.54.1332}{\bibinfo {journal} {Phys. Rev.}}\ }%
  \textbf{\bibinfo {volume} {D54}},\ \bibinfo {pages} {1332} (\bibinfo {year}
  {1996}),\
  \Eprint{http://arxiv.org/abs/astro-ph/9512139}{arXiv:astro-ph/9512139}%
  \bibAnnoteFile{NoStop}{Jungman:1995bz}%
%%CITATION = ASTRO-PH/9512139;%%
\bibitem{Tegmark:1996bz}%
  \BibitemOpen
  \bibfield{author}{%
  \bibinfo {author} {\bibfnamefont{M.}~\bibnamefont{Tegmark}}, \bibinfo
  {author} {\bibfnamefont{A.}~\bibnamefont{Taylor}},\ and\ \bibinfo {author}
  {\bibfnamefont{A.}~\bibnamefont{Heavens}},\ }%
  \bibfield{journal}{%
  \Doi{10.1086/303939}{\bibinfo {journal} {Astrophys. J.}}\ }%
  \textbf{\bibinfo {volume} {480}},\ \bibinfo {pages} {22} (\bibinfo {year}
  {1997}),\
  \Eprint{http://arxiv.org/abs/astro-ph/9603021}{arXiv:astro-ph/9603021}%
  \bibAnnoteFile{NoStop}{Tegmark:1996bz}%
%%CITATION = ASTRO-PH/9603021;%%
\bibitem{Tegmark:1997rp}%
  \BibitemOpen
  \bibfield{author}{%
  \bibinfo {author} {\bibfnamefont{M.}~\bibnamefont{Tegmark}},\ }%
  \bibfield{journal}{%
  \Doi{10.1103/PhysRevLett.79.3806}{\bibinfo {journal} {Phys. Rev. Lett.}}\ }%
  \textbf{\bibinfo {volume} {79}},\ \bibinfo {pages} {3806} (\bibinfo {year}
  {1997}),\
  \Eprint{http://arxiv.org/abs/astro-ph/9706198}{arXiv:astro-ph/9706198
  [astro-ph]}%
  \bibAnnoteFile{NoStop}{Tegmark:1997rp}%
%%CITATION = ASTRO-PH/9706198;%%
\bibitem{Eisenstein:1997ik}%
  \BibitemOpen
  \bibfield{author}{%
  \bibinfo {author} {\bibfnamefont{D.~J.}\ \bibnamefont{Eisenstein}}\ and\
  \bibinfo {author} {\bibfnamefont{W.}~\bibnamefont{Hu}},\ }%
  \bibfield{journal}{%
  \Doi{10.1086/305424}{\bibinfo {journal} {Astrophys. J.}}\ }%
  \textbf{\bibinfo {volume} {496}},\ \bibinfo {pages} {605} (\bibinfo {year}
  {1998}),\
  \Eprint{http://arxiv.org/abs/astro-ph/9709112}{arXiv:astro-ph/9709112}%
  \bibAnnoteFile{NoStop}{Eisenstein:1997ik}%
%%CITATION = ASTRO-PH/9709112;%%
\bibitem{Chang:2007xk}%
  \BibitemOpen
  \bibfield{author}{%
  \bibinfo {author} {\bibfnamefont{T.-C.}\ \bibnamefont{Chang}}, \bibinfo
  {author} {\bibfnamefont{U.-L.}\ \bibnamefont{Pen}}, \bibinfo {author}
  {\bibfnamefont{J.~B.}\ \bibnamefont{Peterson}},\ and\ \bibinfo {author}
  {\bibfnamefont{P.}~\bibnamefont{McDonald}},\ }%
  \bibfield{journal}{%
  \Doi{10.1103/PhysRevLett.100.091303}{\bibinfo {journal} {Phys. Rev. Lett.}}\
  }%
  \textbf{\bibinfo {volume} {100}},\ \bibinfo {pages} {091303} (\bibinfo {year}
  {2008}),\ \Eprint{http://arxiv.org/abs/0709.3672}{arXiv:0709.3672
  [astro-ph]}%
  \bibAnnoteFile{NoStop}{Chang:2007xk}%
%%CITATION = ARXIV:0709.3672;%%
\bibitem{Masui:2009cj}%
  \BibitemOpen
  \bibfield{author}{%
  \bibinfo {author} {\bibfnamefont{K.~W.}\ \bibnamefont{Masui}}, \bibinfo
  {author} {\bibfnamefont{F.}~\bibnamefont{Schmidt}}, \bibinfo {author}
  {\bibfnamefont{U.-L.}\ \bibnamefont{Pen}},\ and\ \bibinfo {author}
  {\bibfnamefont{P.}~\bibnamefont{McDonald}},\ }%
  \bibfield{journal}{%
  \Doi{10.1103/PhysRevD.81.062001}{\bibinfo {journal} {Phys. Rev.}}\ }%
  \textbf{\bibinfo {volume} {D81}},\ \bibinfo {pages} {062001} (\bibinfo {year}
  {2010}),\ \Eprint{http://arxiv.org/abs/0911.3552}{arXiv:0911.3552
  [astro-ph.CO]}%
  \bibAnnoteFile{NoStop}{Masui:2009cj}%
\bibitem{Hall:2012wd}%
  \BibitemOpen
  \bibfield{author}{%
  \bibinfo {author} {\bibfnamefont{A.}~\bibnamefont{Hall}}, \bibinfo {author}
  {\bibfnamefont{C.}~\bibnamefont{Bonvin}},\ and\ \bibinfo {author}
  {\bibfnamefont{A.}~\bibnamefont{Challinor}},\ }%
  \bibfield{journal}{%
  \Doi{10.1103/PhysRevD.87.064026}{\bibinfo {journal} {Phys.Rev.}}\ }%
  \textbf{\bibinfo {volume} {D87}},\ \bibinfo {pages} {064026} (\bibinfo {year}
  {2013}),\ \Eprint{http://arxiv.org/abs/1212.0728}{arXiv:1212.0728
  [astro-ph.CO]}%
  \bibAnnoteFile{NoStop}{Hall:2012wd}%
%%CITATION = ARXIV:1212.0728;%%
\bibitem{Camera:2013kpa}%
  \BibitemOpen
  \bibfield{author}{%
  \bibinfo {author} {\bibfnamefont{S.}~\bibnamefont{Camera}}, \bibinfo {author}
  {\bibfnamefont{M.~G.}\ \bibnamefont{Santos}}, \bibinfo {author}
  {\bibfnamefont{P.~G.}\ \bibnamefont{Ferreira}},\ and\ \bibinfo {author}
  {\bibfnamefont{L.}~\bibnamefont{Ferramacho}},\ }%
  \bibfield{journal}{%
  \Doi{10.1103/PhysRevLett.111.171302}{\bibinfo {journal} {Phys. Rev. Lett.}}\
  }%
  \textbf{\bibinfo {volume} {111}},\ \bibinfo {pages} {171302} (\bibinfo {year}
  {2013}),\ \Eprint{http://arxiv.org/abs/1305.6928}{arXiv:1305.6928
  [astro-ph.CO]}%
  \bibAnnoteFile{NoStop}{Camera:2013kpa}%
%%CITATION = ARXIV:1305.6928;%%
\bibitem{Alcock:1979mp}%
  \BibitemOpen
  \bibfield{author}{%
  \bibinfo {author} {\bibfnamefont{C.}~\bibnamefont{Alcock}}\ and\ \bibinfo
  {author} {\bibfnamefont{B.}~\bibnamefont{Paczynski}},\ }%
  \bibfield{journal}{%
  \bibinfo {journal} {Nature}\ }%
  \textbf{\bibinfo {volume} {281}},\ \bibinfo {pages} {358} (\bibinfo {year}
  {1979})%
  \bibAnnoteFile{NoStop}{Alcock:1979mp}%
%%CITATION = NATUA,281,358;%%
\bibitem{Seo:2003pu}%
  \BibitemOpen
  \bibfield{author}{%
  \bibinfo {author} {\bibfnamefont{H.-J.}\ \bibnamefont{Seo}}\ and\ \bibinfo
  {author} {\bibfnamefont{D.~J.}\ \bibnamefont{Eisenstein}},\ }%
  \bibfield{journal}{%
  \Doi{10.1086/379122}{\bibinfo {journal} {Astrophys. J.}}\ }%
  \textbf{\bibinfo {volume} {598}},\ \bibinfo {pages} {720} (\bibinfo {year}
  {2003}),\
  \Eprint{http://arxiv.org/abs/astro-ph/0307460}{arXiv:astro-ph/0307460
  [astro-ph]}%
  \bibAnnoteFile{NoStop}{Seo:2003pu}%
%%CITATION = ASTRO-PH/0307460;%%
\bibitem{Wang:2006qt}%
  \BibitemOpen
  \bibfield{author}{%
  \bibinfo {author} {\bibfnamefont{Y.}~\bibnamefont{Wang}},\ }%
  \bibfield{journal}{%
  \Doi{10.1086/505384}{\bibinfo {journal} {Astrophys. J.}}\ }%
  \textbf{\bibinfo {volume} {647}},\ \bibinfo {pages} {1} (\bibinfo {year}
  {2006}),\
  \Eprint{http://arxiv.org/abs/astro-ph/0601163}{arXiv:astro-ph/0601163
  [astro-ph]}%
  \bibAnnoteFile{NoStop}{Wang:2006qt}%
%%CITATION = ASTRO-PH/0601163;%%
\bibitem{Majerotto:2012mf}%
  \BibitemOpen
  \bibfield{author}{%
  \bibinfo {author} {\bibfnamefont{E.}~\bibnamefont{Majerotto}}, \bibinfo
  {author} {\bibfnamefont{L.}~\bibnamefont{Guzzo}}, \bibinfo {author}
  {\bibfnamefont{L.}~\bibnamefont{Samushia}}, \bibinfo {author}
  {\bibfnamefont{W.~J.}\ \bibnamefont{Percival}}, \bibinfo {author}
  {\bibfnamefont{Y.}~\bibnamefont{Wang}}, \emph{et~al.},\ }%
  \bibfield{journal}{%
  \Doi{10.1111/j.1365-2966.2012.21323.x}{\bibinfo {journal} {Mon. Not. Roy.
  Astron. Soc.}}\ }%
  \textbf{\bibinfo {volume} {424}},\ \bibinfo {pages} {1392} (\bibinfo {year}
  {2012}),\ \Eprint{http://arxiv.org/abs/1205.6215}{arXiv:1205.6215
  [astro-ph.CO]}%
  \bibAnnoteFile{NoStop}{Majerotto:2012mf}%
%%CITATION = ARXIV:1205.6215;%%
\bibitem{Wang:2010gq}%
  \BibitemOpen
  \bibfield{author}{%
  \bibinfo {author} {\bibfnamefont{Y.}~\bibnamefont{Wang}}, \bibinfo {author}
  {\bibfnamefont{W.}~\bibnamefont{Percival}}, \bibinfo {author}
  {\bibfnamefont{A.}~\bibnamefont{Cimatti}}, \bibinfo {author}
  {\bibfnamefont{P.}~\bibnamefont{Mukherjee}}, \bibinfo {author}
  {\bibfnamefont{L.}~\bibnamefont{Guzzo}}, \emph{et~al.},\ }%
  \bibfield{journal}{%
  \Doi{10.1111/j.1365-2966.2010.17335.x}{\bibinfo {journal} {Mon. Not. Roy.
  Astron. Soc.}}\ }%
  \textbf{\bibinfo {volume} {409}},\ \bibinfo {pages} {737} (\bibinfo {year}
  {2010}),\ \Eprint{http://arxiv.org/abs/1006.3517}{arXiv:1006.3517
  [astro-ph.CO]}%
  \bibAnnoteFile{NoStop}{Wang:2010gq}%
%%CITATION = ARXIV:1006.3517;%%
\bibitem{Camera:2012sf}%
  \BibitemOpen
  \bibfield{author}{%
  \bibinfo {author} {\bibfnamefont{S.}~\bibnamefont{Camera}}, \bibinfo {author}
  {\bibfnamefont{C.}~\bibnamefont{Carbone}},\ and\ \bibinfo {author}
  {\bibfnamefont{L.}~\bibnamefont{Moscardini}},\ }%
  \bibfield{journal}{%
  \Doi{10.1088/1475-7516/2012/03/039}{\bibinfo {journal} {JCAP}}\ }%
  \textbf{\bibinfo {volume} {1203}},\ \bibinfo {pages} {039} (\bibinfo {year}
  {2012}),\ \Eprint{http://arxiv.org/abs/1202.0353}{arXiv:1202.0353
  [astro-ph.CO]}%
  \bibAnnoteFile{NoStop}{Camera:2012sf}%
%%CITATION = ARXIV:1202.0353;%%
\bibitem{1953ApJ...117..134L}%
  \BibitemOpen
  \bibfield{author}{%
  \bibinfo {author} {\bibfnamefont{D.~N.}\ \bibnamefont{{Limber}}},\ }%
  \bibfield{journal}{%
  \Doi{10.1086/145672}{\bibinfo {journal} {Astrophys. J.}}\ }%
  \textbf{\bibinfo {volume} {117}},\ \bibinfo {pages} {134} (\bibinfo {month}
  {Jan.}\ \bibinfo {year} {1953})%
  \bibAnnoteFile{NoStop}{1953ApJ...117..134L}%
\bibitem{Kaiser:1987qv}%
  \BibitemOpen
  \bibfield{author}{%
  \bibinfo {author} {\bibfnamefont{N.}~\bibnamefont{Kaiser}},\ }%
  \bibfield{journal}{%
  \bibinfo {journal} {Mon. Not. Roy. Astron. Soc.}\ }%
  \textbf{\bibinfo {volume} {227}},\ \bibinfo {pages} {1} (\bibinfo {year}
  {1987})%
  \bibAnnoteFile{NoStop}{Kaiser:1987qv}%
%%CITATION = MNRAA,227,1;%%
\bibitem{Hu:2000ee}%
  \BibitemOpen
  \bibfield{author}{%
  \bibinfo {author} {\bibfnamefont{W.}~\bibnamefont{Hu}},\ }%
  \bibfield{journal}{%
  \Doi{10.1103/PhysRevD.62.043007}{\bibinfo {journal} {Phys. Rev.}}\ }%
  \textbf{\bibinfo {volume} {D62}},\ \bibinfo {pages} {043007} (\bibinfo {year}
  {2000}),\
  \Eprint{http://arxiv.org/abs/astro-ph/0001303}{arXiv:astro-ph/0001303}%
  \bibAnnoteFile{NoStop}{Hu:2000ee}%
%%CITATION = ASTRO-PH/0001303;%%
\bibitem{Li:2011wu}%
  \BibitemOpen
  \bibfield{author}{%
  \bibinfo {author} {\bibfnamefont{B.}~\bibnamefont{Li}}, \bibinfo {author}
  {\bibfnamefont{T.~P.}\ \bibnamefont{Sotiriou}},\ and\ \bibinfo {author}
  {\bibfnamefont{J.~D.}\ \bibnamefont{Barrow}},\ }%
  \bibfield{journal}{%
  \Doi{10.1103/PhysRevD.83.104017}{\bibinfo {journal} {Phys. Rev.}}\ }%
  \textbf{\bibinfo {volume} {D83}},\ \bibinfo {pages} {104017} (\bibinfo {year}
  {2011}),\ \Eprint{http://arxiv.org/abs/1103.2786}{arXiv:1103.2786
  [astro-ph.CO]}%
  \bibAnnoteFile{NoStop}{Li:2011wu}%
%%CITATION = ARXIV:1103.2786;%%
\bibitem{Hu:1999ek}%
  \BibitemOpen
  \bibfield{author}{%
  \bibinfo {author} {\bibfnamefont{W.}~\bibnamefont{Hu}},\ }%
  \bibfield{journal}{%
  \bibinfo {journal} {Astrophys. J.}\ }%
  \textbf{\bibinfo {volume} {522}},\ \bibinfo {pages} {L21} (\bibinfo {year}
  {1999}),\
  \Eprint{http://arxiv.org/abs/astro-ph/9904153}{arXiv:astro-ph/9904153}%
  \bibAnnoteFile{NoStop}{Hu:1999ek}%
%%CITATION = ASTRO-PH/9904153;%%
\bibitem{Bartelmann:1999yn}%
  \BibitemOpen
  \bibfield{author}{%
  \bibinfo {author} {\bibfnamefont{M.}~\bibnamefont{Bartelmann}}\ and\ \bibinfo
  {author} {\bibfnamefont{P.}~\bibnamefont{Schneider}},\ }%
  \bibfield{journal}{%
  \Doi{10.1016/S0370-1573(00)00082-X}{\bibinfo {journal} {Phys. Rept.}}\ }%
  \textbf{\bibinfo {volume} {340}},\ \bibinfo {pages} {291} (\bibinfo {year}
  {2001}),\
  \Eprint{http://arxiv.org/abs/astro-ph/9912508}{arXiv:astro-ph/9912508}%
  \bibAnnoteFile{NoStop}{Bartelmann:1999yn}%
%%CITATION = ASTRO-PH/9912508;%%
\bibitem{Kaiser:1996tp}%
  \BibitemOpen
  \bibfield{author}{%
  \bibinfo {author} {\bibfnamefont{N.}~\bibnamefont{Kaiser}},\ }%
  \bibfield{journal}{%
  \Doi{10.1086/305515}{\bibinfo {journal} {Astrophys. J.}}\ }%
  \textbf{\bibinfo {volume} {498}},\ \bibinfo {pages} {26} (\bibinfo {year}
  {1998}),\
  \Eprint{http://arxiv.org/abs/astro-ph/9610120}{arXiv:astro-ph/9610120}%
  \bibAnnoteFile{NoStop}{Kaiser:1996tp}%
%%CITATION = ASTRO-PH/9610120;%%
\bibitem{Wu:2012hs}%
  \BibitemOpen
  \bibfield{author}{%
  \bibinfo {author} {\bibfnamefont{Y.-P.}\ \bibnamefont{Wu}}\ and\ \bibinfo
  {author} {\bibfnamefont{C.-Q.}\ \bibnamefont{Geng}},\ }%
  \bibfield{journal}{%
  \Doi{10.1007/JHEP11(2012)142}{\bibinfo {journal} {JHEP}}\ }%
  \textbf{\bibinfo {volume} {1211}},\ \bibinfo {pages} {142} (\bibinfo {year}
  {2012}),\ \Eprint{http://arxiv.org/abs/1211.1778}{arXiv:1211.1778 [gr-qc]}%
  \bibAnnoteFile{NoStop}{Wu:2012hs}%
%%CITATION = ARXIV:1211.1778;%%
\bibitem{Schimd:2004nq}%
  \BibitemOpen
  \bibfield{author}{%
  \bibinfo {author} {\bibfnamefont{C.}~\bibnamefont{Schimd}}, \bibinfo {author}
  {\bibfnamefont{J.-P.}\ \bibnamefont{Uzan}},\ and\ \bibinfo {author}
  {\bibfnamefont{A.}~\bibnamefont{Riazuelo}},\ }%
  \bibfield{journal}{%
  \Doi{10.1103/PhysRevD.71.083512}{\bibinfo {journal} {Phys. Rev.}}\ }%
  \textbf{\bibinfo {volume} {D71}},\ \bibinfo {pages} {083512} (\bibinfo {year}
  {2005}),\
  \Eprint{http://arxiv.org/abs/astro-ph/0412120}{arXiv:astro-ph/0412120
  [astro-ph]}%
  \bibAnnoteFile{NoStop}{Schimd:2004nq}%
%%CITATION = ASTRO-PH/0412120;%%
\bibitem{Tsujikawa:2008in}%
  \BibitemOpen
  \bibfield{author}{%
  \bibinfo {author} {\bibfnamefont{S.}~\bibnamefont{Tsujikawa}}\ and\ \bibinfo
  {author} {\bibfnamefont{T.}~\bibnamefont{Tatekawa}},\ }%
  \bibfield{journal}{%
  \Doi{10.1016/j.physletb.2008.06.052}{\bibinfo {journal} {Phys.Lett.}}\ }%
  \textbf{\bibinfo {volume} {B665}},\ \bibinfo {pages} {325} (\bibinfo {year}
  {2008}),\ \bibinfo {note} {* Brief entry *},\
  \Eprint{http://arxiv.org/abs/0804.4343}{arXiv:0804.4343 [astro-ph]}%
  \bibAnnoteFile{NoStop}{Tsujikawa:2008in}%
\bibitem{Camera:2011mg}%
  \BibitemOpen
  \bibfield{author}{%
  \bibinfo {author} {\bibfnamefont{S.}~\bibnamefont{Camera}}, \bibinfo {author}
  {\bibfnamefont{A.}~\bibnamefont{Diaferio}},\ and\ \bibinfo {author}
  {\bibfnamefont{V.~F.}\ \bibnamefont{Cardone}},\ }%
  \bibfield{journal}{%
  \Doi{10.1088/1475-7516/2011/01/029}{\bibinfo {journal} {JCAP}}\ }%
  \textbf{\bibinfo {volume} {1101}},\ \bibinfo {pages} {029} (\bibinfo {year}
  {2011}),\ \Eprint{http://arxiv.org/abs/1101.2560}{arXiv:1101.2560
  [astro-ph.CO]}%
  \bibAnnoteFile{NoStop}{Camera:2011mg}%
\bibitem{2010MNRAS.402.1330G}%
  \BibitemOpen
  \bibfield{author}{%
  \bibinfo {author} {\bibfnamefont{J.~E.}\ \bibnamefont{{Geach}}}, \bibinfo
  {author} {\bibfnamefont{A.}~\bibnamefont{{Cimatti}}}, \bibinfo {author}
  {\bibfnamefont{W.}~\bibnamefont{{Percival}}}, \bibinfo {author}
  {\bibfnamefont{Y.}~\bibnamefont{{Wang}}}, \bibinfo {author}
  {\bibfnamefont{L.}~\bibnamefont{{Guzzo}}}, \bibinfo {author}
  {\bibfnamefont{G.}~\bibnamefont{{Zamorani}}}, \bibinfo {author}
  {\bibfnamefont{P.}~\bibnamefont{{Rosati}}}, \bibinfo {author}
  {\bibfnamefont{L.}~\bibnamefont{{Pozzetti}}}, \bibinfo {author}
  {\bibfnamefont{A.}~\bibnamefont{{Orsi}}}, \bibinfo {author}
  {\bibfnamefont{C.~M.}\ \bibnamefont{{Baugh}}}, \bibinfo {author}
  {\bibfnamefont{C.~G.}\ \bibnamefont{{Lacey}}}, \bibinfo {author}
  {\bibfnamefont{B.}~\bibnamefont{{Garilli}}}, \bibinfo {author}
  {\bibfnamefont{P.}~\bibnamefont{{Franzetti}}}, \bibinfo {author}
  {\bibfnamefont{J.~R.}\ \bibnamefont{{Walsh}}},\ and\ \bibinfo {author}
  {\bibfnamefont{M.}~\bibnamefont{{K{\"u}mmel}}},\ }%
  \bibfield{journal}{%
  \Doi{10.1111/j.1365-2966.2009.15977.x}{\bibinfo {journal} {Mon. Not. Roy.
  Astron. Soc.}}\ }%
  \textbf{\bibinfo {volume} {402}},\ \bibinfo {pages} {1330} (\bibinfo {month}
  {Feb.}\ \bibinfo {year} {2010}),\
  \Eprint{http://arxiv.org/abs/0911.0686}{arXiv:0911.0686 [astro-ph.CO]}%
  \bibAnnoteFile{NoStop}{2010MNRAS.402.1330G}%
\bibitem{2010PASP..122..827G}%
  \BibitemOpen
  \bibfield{author}{%
  \bibinfo {author} {\bibfnamefont{B.}~\bibnamefont{{Garilli}}}, \bibinfo
  {author} {\bibfnamefont{M.}~\bibnamefont{{Fumana}}}, \bibinfo {author}
  {\bibfnamefont{P.}~\bibnamefont{{Franzetti}}}, \bibinfo {author}
  {\bibfnamefont{L.}~\bibnamefont{{Paioro}}}, \bibinfo {author}
  {\bibfnamefont{M.}~\bibnamefont{{Scodeggio}}}, \bibinfo {author}
  {\bibfnamefont{O.}~\bibnamefont{{Le F{\`e}vre}}}, \bibinfo {author}
  {\bibfnamefont{S.}~\bibnamefont{{Paltani}}},\ and\ \bibinfo {author}
  {\bibfnamefont{R.}~\bibnamefont{{Scaramella}}},\ }%
  \bibfield{journal}{%
  \Doi{10.1086/654903}{\bibinfo {journal} {PASP}}\ }%
  \textbf{\bibinfo {volume} {122}},\ \bibinfo {pages} {827} (\bibinfo {month}
  {Jul.}\ \bibinfo {year} {2010}),\
  \Eprint{http://arxiv.org/abs/1005.2825}{arXiv:1005.2825 [astro-ph.IM]}%
  \bibAnnoteFile{NoStop}{2010PASP..122..827G}%
\bibitem{1994MNRAS.270..245S}%
  \BibitemOpen
  \bibfield{author}{%
  \bibinfo {author} {\bibfnamefont{I.}~\bibnamefont{{Smail}}}, \bibinfo
  {author} {\bibfnamefont{R.~S.}\ \bibnamefont{{Ellis}}},\ and\ \bibinfo
  {author} {\bibfnamefont{M.~J.}\ \bibnamefont{{Fitchett}}},\ }%
  \bibfield{journal}{%
  \bibinfo {journal} {Mon. Not. Roy. Astron. Soc.}\ }%
  \textbf{\bibinfo {volume} {270}},\ \bibinfo {pages} {245} (\bibinfo {month}
  {Sep.}\ \bibinfo {year} {1994}),\
  \Eprint{http://arxiv.org/abs/arXiv:astro-ph/9402048}{arXiv:astro-ph/9402048}%
  \bibAnnoteFile{NoStop}{1994MNRAS.270..245S}%
\bibitem{2010MNRAS.405.1006O}%
  \BibitemOpen
  \bibfield{author}{%
  \bibinfo {author} {\bibfnamefont{A.}~\bibnamefont{{Orsi}}}, \bibinfo {author}
  {\bibfnamefont{C.~M.}\ \bibnamefont{{Baugh}}}, \bibinfo {author}
  {\bibfnamefont{C.~G.}\ \bibnamefont{{Lacey}}}, \bibinfo {author}
  {\bibfnamefont{A.}~\bibnamefont{{Cimatti}}}, \bibinfo {author}
  {\bibfnamefont{Y.}~\bibnamefont{{Wang}}},\ and\ \bibinfo {author}
  {\bibfnamefont{G.}~\bibnamefont{{Zamorani}}},\ }%
  \bibfield{journal}{%
  \Doi{10.1111/j.1365-2966.2010.16585.x}{\bibinfo {journal} {Mon. Not. Roy.
  Astron. Soc.}}\ }%
  \textbf{\bibinfo {volume} {405}},\ \bibinfo {pages} {1006} (\bibinfo {month}
  {Jun.}\ \bibinfo {year} {2010}),\
  \Eprint{http://arxiv.org/abs/0911.0669}{arXiv:0911.0669 [astro-ph.CO]}%
  \bibAnnoteFile{NoStop}{2010MNRAS.405.1006O}%
\bibitem{LoVerde:2008re}%
  \BibitemOpen
  \bibfield{author}{%
  \bibinfo {author} {\bibfnamefont{M.}~\bibnamefont{LoVerde}}\ and\ \bibinfo
  {author} {\bibfnamefont{N.}~\bibnamefont{Afshordi}},\ }%
  \bibfield{journal}{%
  \Doi{10.1103/PhysRevD.78.123506}{\bibinfo {journal} {Phys. Rev.}}\ }%
  \textbf{\bibinfo {volume} {D78}},\ \bibinfo {pages} {123506} (\bibinfo {year}
  {2008}),\ \Eprint{http://arxiv.org/abs/0809.5112}{arXiv:0809.5112
  [astro-ph]}%
  \bibAnnoteFile{NoStop}{LoVerde:2008re}%
%%CITATION = ARXIV:0809.5112;%%
\bibitem{White:2004kv}%
  \BibitemOpen
  \bibfield{author}{%
  \bibinfo {author} {\bibfnamefont{M.~J.}\ \bibnamefont{White},
  \bibfnamefont{1}},\ }%
  \bibfield{journal}{%
  \Doi{10.1016/j.astropartphys.2004.06.001}{\bibinfo {journal} {Astropart.
  Phys.}}\ }%
  \textbf{\bibinfo {volume} {22}},\ \bibinfo {pages} {211} (\bibinfo {year}
  {2004}),\
  \Eprint{http://arxiv.org/abs/astro-ph/0405593}{arXiv:astro-ph/0405593}%
  \bibAnnoteFile{NoStop}{White:2004kv}%
%%CITATION = ASTRO-PH/0405593;%%
\bibitem{Zentner:2012mv}%
  \BibitemOpen
  \bibfield{author}{%
  \bibinfo {author} {\bibfnamefont{A.~R.}\ \bibnamefont{Zentner}}, \bibinfo
  {author} {\bibfnamefont{E.}~\bibnamefont{Semboloni}}, \bibinfo {author}
  {\bibfnamefont{S.}~\bibnamefont{Dodelson}}, \bibinfo {author}
  {\bibfnamefont{T.}~\bibnamefont{Eifler}}, \bibinfo {author}
  {\bibfnamefont{E.}~\bibnamefont{Krause}}, \emph{et~al.}}%
   (\bibinfo {year} {2012}),\
  \Eprint{http://arxiv.org/abs/1212.1177}{arXiv:1212.1177 [astro-ph.CO]}%
  \bibAnnoteFile{NoStop}{Zentner:2012mv}%
%%CITATION = ARXIV:1212.1177;%%
\bibitem{Giannantonio:2011ya}%
  \BibitemOpen
  \bibfield{author}{%
  \bibinfo {author} {\bibfnamefont{T.}~\bibnamefont{Giannantonio}}, \bibinfo
  {author} {\bibfnamefont{C.}~\bibnamefont{Porciani}}, \bibinfo {author}
  {\bibfnamefont{J.}~\bibnamefont{Carron}}, \bibinfo {author}
  {\bibfnamefont{A.}~\bibnamefont{Amara}},\ and\ \bibinfo {author}
  {\bibfnamefont{A.}~\bibnamefont{Pillepich}},\ }%
  \bibfield{journal}{%
  \Doi{10.1111/j.1365-2966.2012.20604.x}{\bibinfo {journal} {Mon. Not. Roy.
  Astron. Soc.}}\ }%
  \textbf{\bibinfo {volume} {422}},\ \bibinfo {pages} {2854} (\bibinfo {year}
  {2012}),\ \Eprint{http://arxiv.org/abs/1109.0958}{arXiv:1109.0958
  [astro-ph.CO]}%
  \bibAnnoteFile{NoStop}{Giannantonio:2011ya}%
%%CITATION = ARXIV:1109.0958;%%
\bibitem{Hu:2003pt}%
  \BibitemOpen
  \bibfield{author}{%
  \bibinfo {author} {\bibfnamefont{W.}~\bibnamefont{Hu}}\ and\ \bibinfo
  {author} {\bibfnamefont{B.}~\bibnamefont{Jain}},\ }%
  \bibfield{journal}{%
  \Doi{10.1103/PhysRevD.70.043009}{\bibinfo {journal} {Phys. Rev.}}\ }%
  \textbf{\bibinfo {volume} {D70}},\ \bibinfo {pages} {043009} (\bibinfo {year}
  {2004}),\
  \Eprint{http://arxiv.org/abs/astro-ph/0312395}{arXiv:astro-ph/0312395
  [astro-ph]}%
  \bibAnnoteFile{NoStop}{Hu:2003pt}%
%%CITATION = ASTRO-PH/0312395;%%
\bibitem{Zhan:2006gi}%
  \BibitemOpen
  \bibfield{author}{%
  \bibinfo {author} {\bibfnamefont{H.}~\bibnamefont{Zhan}},\ }%
  \bibfield{journal}{%
  \Doi{10.1088/1475-7516/2006/08/008}{\bibinfo {journal} {JCAP}}\ }%
  \textbf{\bibinfo {volume} {0608}},\ \bibinfo {pages} {008} (\bibinfo {year}
  {2006}),\
  \Eprint{http://arxiv.org/abs/astro-ph/0605696}{arXiv:astro-ph/0605696
  [astro-ph]}%
  \bibAnnoteFile{NoStop}{Zhan:2006gi}%
%%CITATION = ASTRO-PH/0605696;%%
\bibitem{Weinberg:2012es}%
  \BibitemOpen
  \bibfield{author}{%
  \bibinfo {author} {\bibfnamefont{D.~H.}\ \bibnamefont{Weinberg}}, \bibinfo
  {author} {\bibfnamefont{M.~J.}\ \bibnamefont{Mortonson}}, \bibinfo {author}
  {\bibfnamefont{D.~J.}\ \bibnamefont{Eisenstein}}, \bibinfo {author}
  {\bibfnamefont{C.}~\bibnamefont{Hirata}}, \bibinfo {author}
  {\bibfnamefont{A.~G.}\ \bibnamefont{Riess}}, \emph{et~al.},\ }%
  \bibfield{journal}{%
  \bibinfo {journal} {Phys. Rept.}}%
   (\bibinfo {year} {2012}),\ \doi{\bibinfo {doi}
  {10.1016/j.physrep/2013.05.001}},\
  \Eprint{http://arxiv.org/abs/1201.2434}{arXiv:1201.2434 [astro-ph.CO]}%
  \bibAnnoteFile{NoStop}{Weinberg:2012es}%
%%CITATION = ARXIV:1201.2434;%%
\bibitem{Trotta:2008qt}%
  \BibitemOpen
  \bibfield{author}{%
  \bibinfo {author} {\bibfnamefont{R.}~\bibnamefont{Trotta}},\ }%
  \bibfield{journal}{%
  \Doi{10.1080/00107510802066753}{\bibinfo {journal} {Contemp. Phys.}}\ }%
  \textbf{\bibinfo {volume} {49}},\ \bibinfo {pages} {71} (\bibinfo {year}
  {2008}),\ \Eprint{http://arxiv.org/abs/0803.4089}{arXiv:0803.4089
  [astro-ph]}%
  \bibAnnoteFile{NoStop}{Trotta:2008qt}%
\bibitem{Taylor:2006aw}%
  \BibitemOpen
  \bibfield{author}{%
  \bibinfo {author} {\bibfnamefont{A.~N.}\ \bibnamefont{Taylor}}, \bibinfo
  {author} {\bibfnamefont{T.~D.}\ \bibnamefont{Kitching}}, \bibinfo {author}
  {\bibfnamefont{D.~J.}\ \bibnamefont{Bacon}},\ and\ \bibinfo {author}
  {\bibfnamefont{A.~F.}\ \bibnamefont{Heavens}},\ }%
  \bibfield{journal}{%
  \Doi{10.1111/j.1365-2966.2006.11257.x}{\bibinfo {journal} {Mon. Not. Roy.
  Astron. Soc.}}\ }%
  \textbf{\bibinfo {volume} {374}},\ \bibinfo {pages} {1377} (\bibinfo {year}
  {2007}),\
  \Eprint{http://arxiv.org/abs/astro-ph/0606416}{arXiv:astro-ph/0606416}%
  \bibAnnoteFile{NoStop}{Taylor:2006aw}%
%%CITATION = ASTRO-PH/0606416;%%
\bibitem{Jeffreys:1961}%
  \BibitemOpen
  \bibfield{author}{%
  \bibinfo {author} {\bibfnamefont{H.}~\bibnamefont{Jeffreys}},\ }%
  \emph{\bibinfo {title} {Theory of Probability}}\ (\bibinfo {year} {1961})\
  \bibinfo {note} {~Oxford, UK: Univ. Pr. (1961) 421 p}%
  \bibAnnoteFile{NoStop}{Jeffreys:1961}%
\end{thebibliography}%

\end{document}